\documentclass[aps,prd,twocolumn,superscriptaddress,nofootinbib,longbibliography]{revtex4-1}

\usepackage{amsmath,amssymb,amsthm,bm,graphicx,xcolor}
\usepackage[colorlinks=true,linkcolor=blue,citecolor=blue,urlcolor=blue]{hyperref}

\begin{document}

\title{Boson star-black hole binaries: initial data and head-on collisions}

\author{Zhuan Ning}
\email{Corresponding author: ningzhuan17@mails.ucas.ac.cn}
\affiliation{School of Fundamental Physics and Mathematical Sciences, Hangzhou Institute for Advanced Study (HIAS), University of Chinese Academy of Sciences (UCAS), Hangzhou 310024, China}
\affiliation{Institute of Theoretical Physics, Chinese Academy of Sciences (CAS), Beijing 100190, China}
\affiliation{University of Chinese Academy of Sciences (UCAS), Beijing 100049, China}

\begin{abstract}
    We present a numerical-relativity study of comparable-mass boson star-black hole (BS-BH) head-on collisions, focusing on both initial-data construction and gravitational-wave (GW) phenomenology. We show that plain superposition can strongly perturb the BS core, leading to large constraint violations and unphysical radial oscillations. To remedy this problem, we introduce a one-body conformal-factor correction and find that it substantially suppresses these artifacts. Using the improved initial data, we analyze GW emission from equal- and unequal-mass BS-BH binaries and compare with matched BS-BS and BH-BH baselines. For equal masses, the BS-BH radiated energy increases with BS compactness and approaches the BH-BH limit for highly compact stars. For unequal masses, the $q = 2$ BS-BH configuration in our sample radiates slightly more GW energy than its matched BH-BH counterpart.
\end{abstract}

\maketitle

\section{Introduction} \label{sec:introduction}

The existence of dark matter (DM) is supported by a broad range of gravitational observations~\cite{Navarro:1995iw,Clowe:2006eq}, while its microphysical nature remains unknown and constitutes one of the most important open problems in modern cosmology and fundamental physics~\cite{Bertone:2004pz,Feng:2010gw}. Despite extensive laboratory and astrophysical searches, no conclusive non-gravitational detection channel has yet been established~\cite{Feng:2010gw,Klasen:2015uma,Kahlhoefer:2017dnp,Cebrian:2022brv}. In this context, gravitational-wave (GW) astronomy~\cite{LIGOScientific:2018mvr,KAGRA:2021vkt,LIGOScientific:2025slb} offers a complementary and potentially unique avenue~\cite{Brito:2017zvb,Bertone:2019irm}: if some fraction of DM can form compact self-gravitating structures, their mergers may leave directly observable imprints in the strong-gravity regime.

Several DM candidates have been proposed, ranging from particle-like candidates such as weakly interacting massive particles (WIMPs)~\cite{Pospelov:2007mp,Arcadi:2017kky,Roszkowski:2017nbc,Schumann:2019eaa} and sterile neutrinos~\cite{Dodelson:1993je,Shi:1998km,Abazajian:2012ys}, to wave-like candidates such as ultralight scalar fields~\cite{Hu:2000ke,Hui:2016ltb,Hui:2021tkt}, and even macroscopic objects such as primordial black holes (PBHs)~\cite{Carr:2016drx,Green:2020jor,Ning:2025ogq,Zeng:2025law,Ning:2025yvj,Carr:2026hot,Ning:2026nfs,Klipfel:2026nzx,Jia:2026psi}. Among these candidates, scalar fields are particularly well motivated and have been studied extensively. They arise naturally in a wide range of theoretical scenarios, including the QCD axion~\cite{Dine:1981rt,Dine:1982ah,Preskill:1982cy}, axion-like particles~\cite{Marsh:2015xka} from string-theory compactifications~\cite{Svrcek:2006yi,Arvanitaki:2009fg}, and more general scalar-field DM models. If endowed with mass, scalar fields minimally coupled to gravity can form macroscopic self-gravitating structures~\cite{Kaup:1968zz,Ruffini:1969qy,Seidel:1991zh}. Complex scalar fields can support static, spherically symmetric configurations in which the field oscillates in time, known as boson stars (BSs)~\cite{Kaup:1968zz,Ruffini:1969qy}, while real scalar fields can instead form long-lived oscillating configurations with a nontrivial time-dependent energy-momentum tensor, called oscillatons~\cite{Seidel:1991zh} (see, e.g., Refs.~\cite{Schunck:2003kk,Liebling:2012fv,Visinelli:2021uve,Bezares:2024btu} for reviews). Both types of solutions can arise naturally as end states of gravitational collapse~\cite{Kaup:1968zz,Ruffini:1969qy,Khlopov:1985fch,Seidel:1991zh,Okawa:2013jba,Chavanis:2016dab,Widdicombe:2018oeo} and have a maximum mass scaling as $M_\mathrm{max} \sim M_\mathrm{pl}^2/\mu$, where $M_\mathrm{pl}$ is the Planck mass and $\mu$ is the scalar-field mass. Consequently, depending on the scalar mass, these objects can span a wide range of masses and play very different cosmological roles.

Ultralight scalar fields with masses in the range $\mu \sim 10^{-22} - 10^{-20}\,\mathrm{eV}$, often referred to as ``fuzzy'' DM, have de Broglie wavelengths on astrophysical scales and can therefore modify halo structure on kiloparsec scales, potentially alleviating some of the small-scale tensions of collisionless cold-DM phenomenology~\cite{Hu:2000ke,Hui:2016ltb,Hui:2021tkt}. The corresponding solitonic halo cores~\cite{Schive:2014dra,Veltmaat:2018dfz} are mathematically related to self-gravitating bosonic configurations, but they are typically much less compact and much larger than relativistic compact stars.

At the same time, axion and axion-like scenarios, as well as string-inspired ``axiverse'' constructions, motivate a much broader mass range of scalar fields. For heavier scalars, such as $\mu \sim 10^{-10} - 10^{-16}\,\mathrm{eV}$, the associated self-gravitating configurations can span masses from solar to supermassive scales. In addition, modifying the scalar self-interaction potential can further enlarge the allowed ranges of mass and compactness, enabling configurations that behave as exotic compact objects (ECOs) or, in some regimes, as black-hole (BH) mimickers~\cite{Cardoso:2014sna,Marks:2025jpt,Staelens:2025wom,Evstafyeva:2025mvx,Wang:2026teu}. The mergers of such objects can produce strong GW signals that may be detectable by current and future GW observatories, and these signals may be distinguishable from standard BH or neutron-star (NS) mergers through their characteristic features.

Motivated by these properties, BS-BS binary mergers have been extensively investigated in numerical relativity, including head-on collisions~\cite{Bernal:2006ci,Palenzuela:2006wp,Choptuik:2009ww,Mundim:2010hi,Cardoso:2016oxy,Helfer:2021brt,Jaramillo:2022zwg,Sanchis-Gual:2022zsr,Evstafyeva:2022bpr,Atteneder:2023pge,Siemonsen:2024snb,Lira:2024cma,Ge:2024itl,Brito:2025rld,Ge:2025btw} and orbital inspirals~\cite{Palenzuela:2007dm,Mundim:2010hi,Palenzuela:2017kcg,Bezares:2017mzk,Bezares:2018qwa,Sanchis-Gual:2020mzb,Bezares:2022obu,Croft:2022bxq,Siemonsen:2023hko,Siemonsen:2023age} for both equal-~\cite{Palenzuela:2006wp,Palenzuela:2007dm,Choptuik:2009ww,Mundim:2010hi,Cardoso:2016oxy,Palenzuela:2017kcg,Bezares:2017mzk,Bezares:2018qwa,Sanchis-Gual:2020mzb,Helfer:2021brt,Jaramillo:2022zwg,Sanchis-Gual:2022zsr,Croft:2022bxq,Siemonsen:2023hko,Atteneder:2023pge,Lira:2024cma,Ge:2024itl,Brito:2025rld,Ge:2025btw} and unequal-mass~\cite{Bernal:2006ci,Bezares:2022obu,Evstafyeva:2022bpr,Siemonsen:2023age,Siemonsen:2024snb} binaries. These studies have revealed rich phenomenology, including waveform degeneracies with BH binaries and the imprint of tidal deformability. Furthermore, other ECO mergers have also been studied, such as oscillaton~\cite{Helfer:2018vtq,Widdicombe:2019woy} and Proca-star (vector cousins of BSs)~\cite{Sanchis-Gual:2018oui,CalderonBustillo:2020fyi,CalderonBustillo:2022cja,Sanchis-Gual:2022mkk} binaries. From an astrophysical perspective, if BSs, fuzzy-DM cores, or other scalar-field configurations populate the Universe, their coexistence with BHs and their inevitable interactions are highly probable. In this context, the nonlinear dynamics of scalar fields coupled to BHs---including scalar-field accretion~\cite{Sanchis-Gual:2014ewa,Hui:2019aqm,Clough:2019jpm,Cardoso:2022nzc,Brito:2015yga,Brito:2015yfh}, scalarization~\cite{Okawa:2014nda,Silva:2017uqg,Zhang:2021nnn,Doneva:2022ewd,Chen:2023eru,Guo:2025flg}, descalarization~\cite{Silva:2020omi,Chen:2022vag}, superradiance~\cite{Brito:2015oca,Baryakhtar:2020gao,Lyu:2025lue}, dynamical friction~\cite{Traykova:2021dua,Boudon:2022dxi}, and dynamical transition processes~\cite{Ning:2023edr,Chen:2023iws,Zhao:2026ykv,Zhao:2026jqp}---have been widely explored in various strongly gravitating systems, revealing rich time-evolution behaviors and phase structures. It is therefore also crucial to understand the dynamics and GW signatures of mixed BS-BH binary mergers for interpreting future observations and constraining the existence of such objects.

Compared with BS-BS binaries, BS-BH binaries remain significantly less explored. Ref.~\cite{Clough:2018exo} presented the first fully relativistic simulations of head-on collisions between oscillatons (axion stars, ASs) and BHs/NSs, finding that in the AS-BH case the remaining scalar cloud after merger can be enhanced by decreasing the compactness of the AS and increasing the BH spin. For BS-BH binaries, Refs.~\cite{Cardoso:2022vpj,Zhong:2023xll} studied piercing collisions, focusing on the large-mass-ratio regime, in which a small BH pierces a much larger BS. They found that the scalar field is almost entirely accreted by the BH for both mini and solitonic BSs. In addition, a small fraction of the scalar field survives outside the BH horizon and forms a long-lived quasi-bound state comoving with the BH, i.e., a ``gravitational atom''.

However, as discussed above, BS masses can span a wide range, and it is plausible that they may be comparable to those of BHs. In this regime, the dynamics of BS-BH mergers can differ significantly from the large-mass-ratio case and may produce distinct GW signatures. In this work, we focus on head-on collisions of comparable-mass BS-BH binaries and study how the mass ratio and binary type (BS-BH versus BS-BS and BH-BH) affect GW emission. Very recently, during the final stage of this work, Ref.~\cite{Marks:2026xvo} appeared. The authors also studied comparable-mass BS-BH collisions, examining GW radiation efficiency across different scalar potentials, BS compactnesses, velocities, and mass ratios. They also presented the first simulation of BS-BH inspirals, showing that an appropriate scalar self-interaction can suppress tidal disruption. In this work, we use a different initial-data construction method and different diagnostics to analyze the merger dynamics and GW signatures. Our results are complementary to those of Ref.~\cite{Marks:2026xvo}, and together they provide a more comprehensive understanding of comparable-mass BS-BH mergers.

From a numerical standpoint, the comparable-mass regime is also one in which the quality of the initial data becomes critical. A direct pointwise superposition of boosted single-object solutions (plain superposition) is simple and widely used~\cite{Cardoso:2022vpj,Zhong:2023xll}, but for BSs it can induce unphysical distortions of the stellar core volume element, increase constraint violations, and trigger spurious premature collapse. This issue has already been identified and cured for BS-BS binaries through metric correction and conformal-factor correction in the equal-~\cite{Helfer:2021brt} and unequal-mass~\cite{Evstafyeva:2022bpr} cases, respectively, leading to significant improvements in initial-data quality. The problem is expected to be even more severe for BS-BH binaries, since the gravitational tail of a BH is much stronger than that of a BS.

In Ref.~\cite{Marks:2026xvo}, the authors mitigated this issue by employing a metric correction method for the BS and a \texttt{TwoPunctures} correction for the BH. In this work, developed independently, we adopt a BS-centered one-body conformal-factor correction derived directly from the unequal-mass BS-BS prescription in the appropriate BS-BH limit. This method is self-consistent and provides a unified framework for constructing initial data for BS-BH and BS-BS binaries on an equal footing, allowing for a more direct comparison of their dynamics and GW signatures. We show that our correction strongly suppresses the initial constraint violations near the BS core and eliminates the spurious premature collapse observed with plain superposition. We then use these improved data to study the merger dynamics and GW emission of BS-BH binaries and compare their behavior with that of BS-BS and BH-BH counterparts.

The paper is organized as follows. In Sec.~\ref{sec:formalism}, we summarize the Einstein-Klein-Gordon model, the $3+1$ formulation, and our numerical infrastructure and diagnostics. Section~\ref{sec:initial_data} presents isolated objects, boost construction, plain superposition, and the improved BS-BH prescription. In Sec.~\ref{sec:results}, we analyze the quality of the initial data and study equal- and unequal-mass head-on collisions. Finally, we conclude in Sec.~\ref{sec:conclusions}. Throughout this work, we use natural units $\hbar = c = G = 1$ and code units with $\mu = 1$ unless stated otherwise. Spacetime indices are denoted by Greek letters ($0\ldots 3$), and spatial indices by Latin letters ($1\ldots 3$).

\section{Formalism} \label{sec:formalism}

\subsection{Model}

The action for a self-gravitating complex scalar field $\varphi$ minimally coupled to gravity is given by
\begin{equation}
    S = \int \mathrm{d}^4x \sqrt{-g} \left[ \frac{R}{16\pi} - \frac{1}{2}\left( g^{\mu\nu}\nabla_{\mu}\bar{\varphi}\nabla_{\nu}\varphi + V(|\varphi|^2) \right) \right],
\end{equation}
where $g$ is the determinant of the spacetime metric $g_{\mu\nu}$, $R$ and $\nabla_{\mu}$ denote the Ricci scalar and the covariant derivative associated with this metric, $\bar{\varphi}$ is the complex conjugate of the scalar field, and $V(|\varphi|^2)$ is the scalar-field potential. Varying the action with respect to the metric and the scalar field yields the Einstein-Klein-Gordon equations:
\begin{align}
    R_{\mu\nu} - \frac{1}{2}R g_{\mu\nu} &= 8\pi T_{\mu\nu}, \\
    \nabla^\mu \nabla_\mu \varphi &= V'\varphi \equiv \frac{dV}{d|\varphi|^2}\varphi,
\end{align}
where the energy-momentum tensor of the complex scalar field is
\begin{equation}
    \begin{aligned}
        T_{\mu\nu} &= \frac{1}{2} \left( \nabla_\mu \bar{\varphi} \nabla_\nu \varphi + \nabla_\nu \bar{\varphi} \nabla_\mu \varphi \right) \\
        & \quad - \frac{1}{2} g_{\mu\nu} \left( g^{\alpha\beta} \nabla_\alpha \bar{\varphi} \nabla_\beta \varphi + V(|\varphi|^2) \right).
    \end{aligned}
\end{equation}
BSs are stationary solutions to the Einstein-Klein-Gordon equations, and their properties and compactness depend crucially on the choice of the potential $V(|\varphi|^2)$. In this work, we focus on the solitonic potential~\cite{Lee:1986tr}:
\begin{equation}
    V = \mu^2 |\varphi|^2 \left( 1 - 2\frac{|\varphi|^2}{\sigma_0^2} \right)^2,
\end{equation}
where $\mu$ denotes the bare mass of the scalar field and $\sigma_0$ regulates the strength of the self-interaction. Note that the non-self-interacting mini BS is recovered in the limit $\sigma_0 \to \infty$. In our numerical implementation, we adopt code units with $\mu = 1$, so all time and length variables are expressed in units of $\mu^{-1}$.

\subsection{$3+1$ formulation and computational infrastructure}

We evolve the Einstein-Klein-Gordon equations using the standard $3+1$ decomposition of spacetime~\cite{Gourgoulhon:2007ue,2008itnr.book.....A,2016nure.book.....S}, which decomposes the spacetime into spatial metric $\gamma_{ij}$, lapse function $\alpha$, and shift vector $\beta^i$ in adapted coordinates $(t,x^i)$. The line element can be written as
\begin{equation}
    ds^2 = -\alpha^2 dt^2 + \gamma_{ij} (dx^i + \beta^i dt)(dx^j + \beta^j dt).
\end{equation}
To cast the Einstein-Klein-Gordon equations into a first-order-in-time system, we introduce the extrinsic curvature
\begin{equation}
    K_{ij} \equiv -\frac{1}{2\alpha} (\partial_t \gamma_{ij} - \mathcal{L}_\beta \gamma_{ij}) = -\frac{1}{2\alpha} (\partial_t \gamma_{ij} - D_i \beta_j - D_j \beta_i),
\end{equation}
and the conjugate momentum of the scalar field
\begin{equation}
    \Pi \equiv -\frac{1}{\alpha} (\partial_t \varphi - \mathcal{L}_\beta \varphi) = -\frac{1}{\alpha} (\partial_t \varphi - \beta^i \partial_i \varphi),
\end{equation}
where $\mathcal{L}_\beta$ denotes the Lie derivative along the shift vector $\beta^i$, and $D_i$ is the covariant derivative compatible with the spatial metric $\gamma_{ij}$. The evolution equations for $\gamma_{ij}$, $K_{ij}$, $\varphi$, and $\Pi$ then follow from the Einstein-Klein-Gordon equations:
\begin{subequations}
    \begin{align}
        \partial_{t} \gamma_{ij} & = - 2 \alpha K_{ij} + \mathcal{L}_{\beta} \gamma_{ij} , \\
        \partial_{t} K_{ij} & = - D_{i} D_{j} \alpha + \alpha \left( \mathcal{R}_{ij} + K K_{ij} - 2 K_{ik} {K^{k}}_{j} \right. \nonumber \\
                            & \left. \quad + 4\pi \left[ (S-\rho) \gamma_{ij} - 2 S_{ij} \right] \right) + \mathcal{L}_{\beta} K_{ij}, \\
        \partial_{t} \varphi & = - \alpha \Pi + \mathcal{L}_{\beta} \varphi, \\
        \partial_{t} \Pi &  = \alpha \left( K \Pi + V' \varphi - D_i D^i \varphi \right) \nonumber \\
                         & \quad - (D_i \alpha) D^i \varphi + \mathcal{L}_{\beta} \Pi .
    \end{align}
\end{subequations}
These equations are subject to the Hamiltonian and momentum constraints:
\begin{align}
\label{eq:Hamiltonian}
    \mathcal{H} & \equiv \mathcal{R} + K^2 - K_{ij} K^{ij} - 16 \pi \rho = 0, \\
    \mathcal{M}_{i} & \equiv D_{j} {K^{j}}_{i} - D_{i} K - 8\pi j_i = 0,
\end{align}
where $K \equiv \gamma^{ij} K_{ij}$, and $\mathcal{R}_{ij}$ and $\mathcal{R}$ denote the Ricci tensor and Ricci scalar associated with the spatial metric $\gamma_{ij}$, respectively. The source terms $\rho$, $j_i$, $S_{ij}$, and $S$ are defined as
\begin{equation}
    \begin{aligned}
        \rho & \equiv n_{a} n_{b} T^{ab} = \frac{1}{2}(\bar{\Pi} \Pi + \partial^a \bar{\varphi}\partial_a\varphi + V),\\
        j_i &\equiv -\gamma_{ia} n_{b} T^{ab} = \frac{1}{2}(\bar{\Pi}\partial_i \varphi + \Pi \partial_i \bar{\varphi}), \\
        S_{ij} &\equiv \gamma_{ia} \gamma_{jb} T^{ab} = \frac{1}{2}(\partial_{i}\bar{\varphi}\partial_{j}\varphi + \partial_{j}\bar{\varphi}\partial_{i}\varphi) \\
        & \quad - \frac{1}{2}\gamma_{ij} \left(\partial^a \bar{\varphi}\partial_a \varphi-\bar{\Pi} \Pi+V\right), \\
        S & \equiv \gamma^{ij} S_{ij} = \frac{1}{2}(3\bar{\Pi}\Pi - \partial^a \bar{\varphi}\partial_a \varphi - 3V).
    \end{aligned}
\end{equation}

For stable numerical evolutions, the evolution system must be written in a strongly hyperbolic form. To this end, we use the publicly available code {\sc GRChombo}~\cite{Clough:2015sqa,Radia:2021smk,Andrade:2021rbd}, which is built on the {\sc Chombo} framework for adaptive mesh refinement (AMR)~\cite{Adams:2015kgr} and evolves the Einstein equations using the Baumgarte-Shapiro-Shibata-Nakamura-Oohara-Kojima (BSSNOK)~\cite{Nakamura:1987zz,Shibata:1995we,Baumgarte:1998te} and covariant conformal Z4 (CCZ4)~\cite{Alic:2011gg, Alic:2013xsa} formulations. In this work, we choose the CCZ4 formulation because of its constraint-damping properties. Specifically, we describe the spacetime in terms of conformally rescaled and trace-split variables,
\begin{equation}
    \begin{aligned}
        \chi &\equiv (\det \gamma_{ij})^{-1/3}, \\
        \tilde{\gamma}_{ij} &\equiv \chi \gamma_{ij}, \\
        \tilde{A}_{ij} &\equiv \chi \left(K_{ij}-\frac{1}{3}\gamma_{ij}K\right), \\
        \tilde{\Gamma}^i &\equiv \tilde{\gamma}^{ab} \tilde{\Gamma}^i_{ab},
    \end{aligned}
\end{equation}
and we choose $\kappa_1 \rightarrow \kappa_1/\alpha$, $\kappa_1 = 0.1$, $\kappa_2 = 0$, and $\kappa_3 = 1$ for the constraint-damping parameters~\cite{Radia:2021smk}. The full Einstein equations in the CCZ4 formulation can be found in Sec.~III F of Ref.~\cite{Alic:2013xsa}. In addition, we use the {\sc ExoZvezda}~\cite{Helfer:2021brt,Croft:2022gks,Croft:2022bxq,Evstafyeva:2022bpr,Evstafyeva:2023kfg,Evstafyeva:2024qvp} code to compute one-dimensional BS solutions in isotropic gauge and to dynamically evolve the complex scalar field.

{\sc GRChombo} discretizes the CCZ4 equations using the method of lines, with fourth-order finite-difference stencils in space and a fourth-order Runge-Kutta integrator in time. Unless otherwise indicated, we choose the Courant-Friedrichs-Lewy (CFL) factor to be $0.25$ and the Kreiss-Oliger dissipation coefficient to be $0.3$. We use the second derivatives of the complex scalar field $\varphi$ and the conformal factor $\chi$ as tagging criteria for AMR, with thresholds of $0.5$ and $0.25$, respectively. We employ a domain of length $512$ with a nested grid hierarchy of $L = 7$ refinement levels, with grid spacing $dx = 1/32$ on the finest level and a factor of $2$ increase between successive levels farther out. For GW extraction, we choose an extraction radius of $R_\mathrm{ex} = 120$. We impose reflective boundary conditions on the $y = 0$ and $z = 0$ planes, since the collision always occurs along the $x$ axis, and radiative (outgoing) boundary conditions on the remaining boundaries. We use the standard $1 + \mathrm{log}$ and Gamma-driver gauge conditions~\cite{2008itnr.book.....A}. Our fiducial Gamma-driver damping is $\eta = 1$. General convergence and wave-zone constraint tests are presented in Appendix~\ref{app:convergence}.

\subsection{Diagnostics}

The $U(1)$ symmetry of the complex scalar-field action gives rise to a conserved current and an associated Noether charge,
\begin{align}
    J^\mu &= \frac{i}{2}(\varphi\nabla^\mu\bar\varphi - \bar\varphi\nabla^\mu \varphi), \\
    N &= \int \mathrm{d}^3 x \sqrt{-g} J^0,
\end{align}
which can be interpreted as a particle number. This diagnostic is particularly useful for tracking the scalar-field content during the merger and for understanding the accretion dynamics onto the BH.

We extract the outgoing GW radiation in terms of the Newman-Penrose scalar $\Psi_4$~\cite{Newman:1961qr,Bishop:2016lgv,Radia:2021smk} and decompose it into spin-weight $s=-2$ spherical harmonics $Y^{-2}_{\ell m}$~\cite{Brugmann:2008zz} according to
\begin{equation}
   \Psi_{4}^{\ell m} (t, R_\mathrm{ex}) = \int_{S^2} \Psi_4(t, R_{\rm ex}, \theta, \phi) \overline{Y^{-2}_{\ell m}} (\theta, \phi) \sin \theta \mathrm{d}\theta \mathrm{d}\phi,
\end{equation}
where $S^2$ is a 2-sphere of fixed extraction radius $R_\mathrm{ex}$. By default, {\sc GRChombo} performs this decomposition with respect to the grid's $z$ axis as the polar axis. However, in our setup the binary undergoes a head-on collision strictly along the $x$ axis, so this choice obscures the underlying axisymmetry of the physical system. To recover the physically relevant axisymmetric modes aligned with the collision axis, we perform a passive rotation of the extracted multipole moments using the Wigner $D$-matrices:
\begin{equation}
    {\Psi_{4}'}^{\ell m'}(t) = \sum_{m=-\ell}^{\ell} D^l_{m' m}(\alpha, \beta, \gamma) \Psi_{4}^{\ell m}(t),
\end{equation}
where ${\Psi_{4}'}^{\ell m'}$ are the modes in the rotated frame and the Euler angles are $(\alpha, \beta, \gamma) = (0, \pi/2, 0)$. In what follows, we drop the prime and refer to the rotated modes simply as $\Psi_{4}^{\ell m}$.

We also compute the cumulative GW energy $E_\mathrm{GW}$ from the Newman-Penrose scalar multipoles $\Psi_{4}^{\ell m}$ via~\cite{Isaacson:1968zza}
\begin{align}
  \dot{E}_\mathrm{GW}^{\ell m}(t) &= \frac{R_\mathrm{ex}^2}{16 \pi} \left| \int_{-\infty}^t \Psi_{4}^{\ell m}(\tilde{t}) \mathrm{d} \tilde{t} \right|^2, \\
  \dot{E}_\mathrm{GW}(t) &= \sum_{\ell, m} \dot{E}_\mathrm{GW}^{\ell m}(t), \\
  E_\mathrm{GW}(t) &= \int_{t_0}^t \dot{E}_\mathrm{GW}(\tilde{t}) \mathrm{d} \tilde{t}.
\end{align}
We begin the time integration at $t_0 = 50$ to exclude any spurious or ``junk'' radiation present in the initial data.

\section{Initial data} \label{sec:initial_data}

\subsection{Isolated boson star and black hole}

To construct binary initial data, we first require stationary solutions for an isolated BS and an isolated BH in their respective rest frames. For convenience in converting to Cartesian coordinates and to avoid coordinate singularities, both compact objects are initially prepared in isotropic coordinates. The static, spherically symmetric metric takes the conformally flat form
\begin{equation}
    \mathrm{d}s^2 = -\alpha^2(r) \mathrm{d}t^2 + \psi^4(r) \delta_{ij} \mathrm{d}x^i \mathrm{d}x^j,
\end{equation}
where $r = \sqrt{x^2+y^2+z^2}$ is the isotropic radial coordinate, $\alpha$ is the lapse function, and $\psi$ is the conformal factor. Since the shift vector vanishes in the rest frame ($\beta^i = 0$), the extrinsic curvature $K_{ij}$ is identically zero.

For the \textit{boson star}, the scalar field is assumed to be stationary and harmonic in time,
\begin{equation}
    \varphi(t,r) = A(r) e^{i\omega t},
\end{equation}
where $A(r)$ is a real radial amplitude and $\omega$ is a constant eigenfrequency. Substituting this ansatz and the isotropic metric into the Einstein-Klein-Gordon equations reduces the system to a set of coupled ordinary differential equations for $\alpha(r)$, $\psi(r)$, and $A(r)$. This system forms an eigenvalue problem, which we solve using a standard shooting method. By specifying the central scalar-field amplitude, $\varphi_{ctr} = A(0)$, the algorithm iteratively adjusts the frequency $\omega$ until the asymptotically flat boundary conditions $A \to 0$ and $\alpha, \psi \to 1$ as $r \to \infty$ are satisfied.

For the \textit{black hole}, we use the standard isotropic Schwarzschild solution (puncture data) for a BH of bare mass $M_\mathrm{BH}$,
\begin{equation}
    \begin{aligned}
        \alpha_{BH}(r) &= \frac{1 - M_\mathrm{BH}/2r}{1 + M_\mathrm{BH}/2r}, \\
        \psi_{BH}(r) &= 1 + \frac{M_\mathrm{BH}}{2r},
    \end{aligned}
\end{equation}
with the scalar field trivially satisfying $\varphi = 0$ and the conjugate momentum $\Pi = 0$. In the rest frame, the Arnowitt-Deser-Misner (ADM)~\cite{Arnowitt:1962hi} mass of the isolated BH is exactly $M_\mathrm{BH}$.

\subsection{Lorentz boost}

To construct a binary system in which the objects collide with prescribed initial velocities, the isolated rest-frame solutions must be boosted to a chosen macroscopic velocity $v$. Without loss of generality, we consider a constant-velocity boost along the $x$-axis. We denote the rest-frame coordinates by $x^\mu = (t, x, y, z)$ and the lab-frame (boosted) coordinates by $\tilde{x}^{\tilde{\alpha}} = (\tilde{t}, \tilde{x}, \tilde{y}, \tilde{z})$. The Lorentz transformation is governed by the standard boost matrix with rapidity $\eta \equiv \mathrm{arctanh}(v)$:
\begin{equation}
    \begin{aligned}
        {\Lambda^{\tilde{\alpha}}}_{\mu} &= \begin{pmatrix}
            \cosh\eta & \sinh\eta & 0 & 0 \\
            \sinh\eta & \cosh\eta & 0 & 0 \\
            0 & 0 & 1 & 0 \\
            0 & 0 & 0 & 1
        \end{pmatrix} \\
        \quad \Leftrightarrow \quad
        {\Lambda^{\mu}}_{\tilde{\alpha}} &= \begin{pmatrix}
            \cosh\eta & -\sinh\eta & 0 & 0 \\
            -\sinh\eta & \cosh\eta & 0 & 0 \\
            0 & 0 & 1 & 0 \\
            0 & 0 & 0 & 1
        \end{pmatrix}.
    \end{aligned}
\end{equation}

Applying the coordinate transformation $\tilde{x}^{\tilde{\alpha}} = {\Lambda^{\tilde{\alpha}}}_{\mu} x^{\mu}$ and the corresponding tensor transformation $\tilde{g}_{\tilde{\alpha}\tilde{\beta}} = {\Lambda^{\mu}}_{\tilde{\alpha}} {\Lambda^{\nu}}_{\tilde{\beta}} g_{\mu\nu}$ to the isotropic metric, and evaluating the resulting fields on the initial hypersurface $\tilde{t}=0$, yields the exact $3+1$ ADM variables in the boosted frame. The boosted lapse $\tilde{\alpha}$ and shift vector $\tilde{\beta}^{\tilde{i}}$ are
\begin{align}
    \tilde{\alpha}^2 &= \frac{\psi^4 \alpha^2}{\psi^4 \cosh^2\eta - \alpha^2 \sinh^2\eta}, \\
    \tilde{\beta}^{\tilde{x}} &= \frac{\sinh\eta \cosh\eta (\alpha^2 - \psi^4)}{\psi^4 \cosh^2\eta - \alpha^2 \sinh^2\eta}, \quad \tilde{\beta}^{\tilde{y}} = \tilde{\beta}^{\tilde{z}} = 0.
\end{align}
The boosted spatial metric $\tilde{\gamma}_{ij}$ is no longer conformally flat and becomes
\begin{equation}
    \tilde{\gamma}_{\tilde{i}\tilde{j}} = \begin{pmatrix}
        \psi^4 \cosh^2\eta - \alpha^2 \sinh^2\eta & 0 & 0 \\
        0 & \psi^4 & 0 \\
        0 & 0 & \psi^4
    \end{pmatrix}.
\end{equation}
The extrinsic curvature is computed analytically from $\tilde{K}_{\tilde{i}\tilde{j}} = -\frac{1}{2\tilde{\alpha}} (\partial_{\tilde{t}} - \mathcal{L}_{\tilde{\beta}}) \tilde{\gamma}_{\tilde{i}\tilde{j}}$~\cite{Croft:2022ecn}. The nonzero components are
\begin{equation}
    \begin{aligned}
        \tilde{K}_{\tilde{x}\tilde{x}} &= \tilde{\alpha}\sinh\eta\cosh^2\eta\frac{x}{r}\left(2\frac{\alpha'}{\alpha} - 2\frac{\psi'}{\psi} - v^2\frac{\alpha\alpha'}{\psi^4}\right), \\
        \tilde{K}_{\tilde{x}\tilde{y}} &= \tilde{K}_{\tilde{y}\tilde{x}} = \tilde{\alpha}\sinh\eta\cosh\eta\frac{y}{r}\left(\frac{\alpha'}{\alpha} - 2\frac{\psi'}{\psi}\right), \\
        \tilde{K}_{\tilde{x}\tilde{z}} &= \tilde{K}_{\tilde{z}\tilde{x}} = \tilde{\alpha}\sinh\eta\cosh\eta\frac{z}{r}\left(\frac{\alpha'}{\alpha} - 2\frac{\psi'}{\psi}\right), \\
        \tilde{K}_{\tilde{y}\tilde{y}} &= \tilde{K}_{\tilde{z}\tilde{z}} = 2\tilde{\alpha}\sinh\eta\frac{x}{r}\frac{\psi'}{\psi},
    \end{aligned}
\end{equation}
where $x=\tilde{x}\cosh\eta$, $y=\tilde{y}$, $z=\tilde{z}$, and $r = \sqrt{x^2+y^2+z^2}$. Primes denote derivatives with respect to $r$.

For the BS, the scalar field and its conjugate momentum must also be transformed to the boosted frame. Evaluated on the $\tilde{t}=0$ slice, the scalar field acquires a spatially modulated phase:
\begin{equation}
    \varphi(\tilde{\boldsymbol{x}}) = \varphi(\boldsymbol{x}) = A(r) e^{-i \omega \tilde{x} \sinh\eta}.
\end{equation}
Finally, the boosted conjugate momentum $\tilde{\Pi}(\tilde{\boldsymbol{x}}) = -\frac{1}{\tilde{\alpha}} (\partial_{\tilde{t}} - \tilde{\beta}^{\tilde{i}} \partial_{\tilde{i}}) \varphi(\tilde{\boldsymbol{x}})$ requires evaluating the time derivative in the lab frame. Applying the chain rule through the Lorentz transformation gives
\begin{equation}
    \begin{aligned}
        \tilde{\Pi}(\tilde{\boldsymbol{x}}) &= -\frac{1}{\tilde{\alpha}} \left[ i\omega A (\cosh\eta + \tilde{\beta}^{\tilde{x}}\sinh\eta) \right. \\
        & \quad \left. - \frac{x}{r}A' (\sinh\eta + \tilde{\beta}^{\tilde{x}}\cosh\eta) \right] e^{-i\omega\tilde{x}\sinh\eta}.
    \end{aligned}
\end{equation}
This boost formalism is applied independently to the numerically integrated BS solution and the exact Schwarzschild BH solution (for which $\varphi = 0$ and $\Pi = 0$), preparing both objects for the subsequent superposition procedure. In the following, we focus on the lab-frame variables and drop the tildes for notational simplicity, with the understanding that all quantities now refer to the boosted frame.

\subsection{Plain superposition and its malaise}

The simplest way to construct binary initial data is to superpose the two isolated boosted solutions obtained above pointwise. We assume two compact objects, labeled A and B, initially located at $x_\mathrm{A}^i$ and $x_\mathrm{B}^i$, with initial separation $d = ||x_\mathrm{A}^i - x_\mathrm{B}^i||$. In terms of the $3+1$ ADM variables, the superposition is performed as follows:
\begin{subequations} \label{eq:plain_superposition}
    \begin{align}
        \gamma_{ij} &= \gamma_{ij}^\mathrm{A} + \gamma_{ij}^\mathrm{B} - \delta_{ij}, \label{eq:plain_gamma} \\
        K_{ij} &= \gamma_{m(i} \left[K_{j)n}^\mathrm{A} \gamma_\mathrm{A}^{mn} + K_{j)n}^\mathrm{B} \gamma_\mathrm{B}^{nm} \right], \\
        \varphi &= \varphi_\mathrm{A} + \varphi_\mathrm{B}, \\
        \Pi &= \Pi_\mathrm{A} + \Pi_\mathrm{B}.
    \end{align}
\end{subequations}
Here, the flat metric $\delta_{ij}$ is subtracted to ensure the correct asymptotic behavior at spatial infinity. For a BS-BH binary, we identify A with the BS and B with the BH, so that $\varphi_\mathrm{B} = \Pi_\mathrm{B}=0$. This construction is commonly referred to as the ``plain superposition'' method and has been widely used in numerical-relativity simulations of BH-BH~\cite{Shibata:2008rq,Okawa:2011fv,Sperhake:2019oaw}, BS-BS~\cite{Palenzuela:2006wp,Palenzuela:2007dm,Palenzuela:2017kcg}, and BS-BH~\cite{Cardoso:2022vpj,Zhong:2023xll} systems. For pure BH spacetimes, this approximation has proved remarkably successful. However, because BSs are extended, horizonless objects supported by a delicate balance between scalar-field pressure and gravity, this simple construction can generate severe unphysical artifacts for them~\cite{Helfer:2018vtq,Helfer:2021brt,Evstafyeva:2022bpr}. In particular, BSs are highly vulnerable to superposition artifacts near their centers because they lack the protective character of a BH horizon.

The main issue arises in the expression for the spatial metric $\gamma_{ij}$, Eq.~\eqref{eq:plain_gamma}. At the center of the BS, $x^i = x^i_\mathrm{A}$, the gravitational field of the companion BH has not completely decayed to flat Minkowski space at finite initial separation $d$. As a result, the metric at the BS center is artificially perturbed by
\begin{equation}
  \delta \gamma_{ij} = \gamma_{ij}^\mathrm{B}(x^i_\mathrm{A}) - \delta_{ij},
  \label{eq:metricpert}
\end{equation}
away from the isolated equilibrium value $\gamma_{ij}^\mathrm{A}(x_\mathrm{A}^i)$. Physically, this perturbation distorts the volume element $\sqrt{\gamma}$ at the BS core, effectively injecting spurious mass-energy and compressing the BS away from its delicate equilibrium. In practice, this artificial compression increases the Hamiltonian-constraint violation, induces unphysical radial oscillations, and alters the GW signal. For sufficiently compact models (especially solitonic BSs), it can even trigger premature collapse into a BH before the physical merger begins. This is the principal ``malaise'' of plain superposition in BS-BH initial data: even when the asymptotic behavior is acceptable, the local BS core can already be driven away from equilibrium at $t=0$.

In Refs.~\cite{Cardoso:2022vpj,Zhong:2023xll}, the BH is much smaller than the BS and the initial separation is sufficiently large, so the superposition artifacts are relatively mild and do not significantly affect the dynamics. However, as we move toward more comparable mass ratios and smaller initial separations in order to systematically explore head-on collision dynamics, these artifacts become increasingly severe and can even induce premature collapse. It is therefore crucial to develop an improved superposition method that mitigates these artifacts and provides more accurate initial data for BS-BH binaries. This is the subject of the next subsection.

\subsection{Improved superposition for boson star-black hole binaries} \label{sec:improved_superposition}

For the BS-BS case, Ref.~\cite{Helfer:2021brt} proposed an improved superposition method for equal-mass binaries that effectively removes the superposition artifacts by modifying the metric superposition to
\begin{equation}
    \gamma_{ij} = \gamma_{ij}^\mathrm{A} + \gamma_{ij}^\mathrm{B} - \gamma_{ij}^\mathrm{B}(x_\mathrm{A}^i) = \gamma_{ij}^\mathrm{A} + \gamma_{ij}^\mathrm{B} - \gamma_{ij}^\mathrm{A}(x_\mathrm{B}^i).
\end{equation}
This method was later generalized to unequal-mass binaries in Ref.~\cite{Evstafyeva:2022bpr}. The key idea is to retain the conformal metric of the plain superposition and modify only its conformal factor so that the BS center recovers the isolated-star volume element. The detailed prescription is given in Sec.~4.2 of Ref.~\cite{Evstafyeva:2022bpr}; here we summarize the main steps.

Starting from the plainly superposed metric in Eq.~\eqref{eq:plain_gamma}, we define a conformal factor $\lambda$ by
\begin{equation}
    \tilde{\gamma}_{ij} = \lambda^{-1} \gamma_{ij},
    \quad
    \lambda = \gamma^{1/3},
    \label{eq:conf_metric_bsbh}
\end{equation}
where $\det \tilde{\gamma}_{ij}=1$ by construction, so that $\lambda=\chi^{-1}$. The improved method leaves $\tilde{\gamma}_{ij}$ unchanged and modifies only $\lambda$ to restore the equilibrium volume element at the object centers. For two generic objects A and B, the target conditions are
\begin{align}
    \lambda_\mathrm{new}(x_\mathrm{A}^i)
    &= \lambda(x_\mathrm{A}^i) + \delta\lambda(x_\mathrm{A}^i)
    = \lambda_\mathrm{A}(x_\mathrm{A}^i),
    \label{eq:conditionAB1_bsbh} \\
    \lambda_\mathrm{new}(x_\mathrm{B}^i)
    &= \lambda(x_\mathrm{B}^i) + \delta\lambda(x_\mathrm{B}^i)
    = \lambda_\mathrm{B}(x_\mathrm{B}^i).
    \label{eq:conditionAB2_bsbh}
\end{align}
With weight functions $w_\mathrm{A}(x^i)$ and $w_\mathrm{B}(x^i)$, one introduces the ansatz
\begin{equation}
    \lambda_\mathrm{new}(x^i) = \lambda(x^i) + w_\mathrm{A}(x^i) h_\mathrm{A} + w_\mathrm{B}(x^i) h_\mathrm{B},
    \label{eq:lambdanew_bsbh}
\end{equation}
where the coefficients $h_\mathrm{A}$ and $h_\mathrm{B}$ are determined by the resulting $2\times 2$ linear system,
\begin{subequations}
    \begin{align}
        h_\mathrm{A} &=
        \frac{-w_\mathrm{B}(x_\mathrm{A}^i)\delta \lambda(x_\mathrm{B}^i)
                    + w_\mathrm{B}(x_\mathrm{B}^i)\delta \lambda (x_\mathrm{A}^i)}
                {w_\mathrm{A}(x_\mathrm{A}^i) w_\mathrm{B}(x_\mathrm{B}^i) - w_\mathrm{A}(x_\mathrm{B}^i)w_\mathrm{B}(x_\mathrm{A}^i)}
                \, , \\
        h_\mathrm{B} &=
        \frac{w_\mathrm{A}(x_\mathrm{A}^i)\delta \lambda(x_\mathrm{B}^i)
                    - w_\mathrm{A}(x_\mathrm{B}^i)\delta \lambda (x_\mathrm{A}^i)}
                {w_\mathrm{A}(x_\mathrm{A}^i) w_\mathrm{B}(x_\mathrm{B}^i) - w_\mathrm{A}(x_\mathrm{B}^i)w_\mathrm{B}(x_\mathrm{A}^i)}.
    \end{align}
    \label{eq:hAB_bsbh}
\end{subequations}
The corrected metric is then reconstructed as
\begin{equation}
    \gamma^\mathrm{new}_{ij} = \left(\frac{\lambda_\mathrm{new}}{\lambda}\right)\gamma_{ij}
    = \frac{\lambda_\mathrm{new}}{\gamma^{1/3}}\gamma_{ij},
    \label{eq:gamma_new_bsbh}
\end{equation}
and a convenient asymptotically flat choice for the weight functions is
\begin{equation}
    w_\mathrm{J}(x^i) = \frac{1}{\sqrt{R_\mathrm{J}^2 + r_\mathrm{J}^2}},
    \quad
    r_\mathrm{J} \equiv ||x^i-x^i_\mathrm{J}||,
    \quad
    \mathrm{J}\in\{\mathrm{A},\mathrm{B}\},
    \label{eq:weight_function_bsbh}
\end{equation}
where $R_\mathrm{J}$ is a tunable parameter that controls the spatial extent of the correction.

For our BS-BH problem, we adopt a similar strategy to remove the superposition artifacts at the BS center, with $\mathrm{A} \equiv \mathrm{BS}$ and $\mathrm{B} \equiv \mathrm{BH}$. However, applying this scheme to a heterogeneous BS-BH binary requires a careful treatment of the BH puncture. First, the conformal factor of the puncture BH diverges at its center, $\lambda_{\mathrm{BH}}(x_{\mathrm{BH}}^i) = \psi_{\mathrm{BH}}^4(x_{\mathrm{BH}}^i) \to \infty$ as $r \to 0$, so the relative error in the conformal volume element induced by the smooth BS background vanishes in the continuum limit. Thus, the required correction at the BH center is identically zero: $\delta\lambda_{\mathrm{BH}} = \lambda_{\mathrm{BH}}(x_{\mathrm{BH}}^i) - \lambda(x_{\mathrm{BH}}^i) \to 0$. Second, the BH interior is inherently immune to superposition artifacts because of the presence of the horizon. Therefore, we may choose the BH correction radius to satisfy $R_{\mathrm{BH}} \to 0$, which implies $w_{\mathrm{BH}}(x_{\mathrm{BH}}^i) \to \infty$.

Taking these two exact physical limits, $\delta\lambda_{\mathrm{BH}} = 0$ and $R_{\mathrm{BH}} \to 0$, within the full $2 \times 2$ system, Eq.~\eqref{eq:hAB_bsbh} reduces to a one-body correction centered on the BS,
\begin{equation}
    h_\mathrm{BS} = \frac{\delta\lambda_\mathrm{BS}(x_\mathrm{BS}^i)}{w_\mathrm{BS}(x_\mathrm{BS}^i)},
    \quad
    h_\mathrm{BH}=0,
\end{equation}
where $\delta\lambda_{\mathrm{BS}} = \lambda_{\mathrm{BS}}(x_{\mathrm{BS}}^i) - \lambda(x_{\mathrm{BS}}^i)$. Therefore,
\begin{equation}
    \lambda_\mathrm{new}(x^i)
    = \lambda(x^i)
    + w_\mathrm{BS}(x^i)\frac{\delta\lambda_\mathrm{BS}(x_\mathrm{BS}^i)}{w_\mathrm{BS}(x_\mathrm{BS}^i)}.
    \label{eq:lambda_new_onebody_bsbh}
\end{equation}
The improved spatial metric then follows from Eq.~\eqref{eq:gamma_new_bsbh}. This construction preserves the plain-superposition structure away from the BS core while enforcing the correct central volume element, which is the key ingredient for curing the spurious distortion discussed above. This BS-centered one-body conformal-factor correction is the BS-BH limit of the unequal-mass two-body BS-BS prescription, and in our simulations it significantly reduces the initial constraint violations and removes the spurious premature collapse seen in plain superposition, as we will demonstrate in the next section.

To complete the specification of the initial data for the $3+1$ evolution, we must also prescribe the initial gauge variables, even though their values do not affect the physical content of the initial spatial hypersurface. For the BS-BS case, one could similarly construct a superposition for the lapse $\alpha$ and shift vector $\beta^i$ as in Eq.~\eqref{eq:plain_superposition}. However, to stably evolve the BS-BH and BH-BH binaries in the moving-puncture framework, we instead initialize them using the standard pre-collapsed prescription:
\begin{equation}
    \alpha = \sqrt{\chi} = \sqrt{\lambda_\mathrm{new}^{-1}}, \quad \beta^i = 0.
\end{equation}

Although residual constraint violations remain in the initial data because of the superposition procedure, the improved method substantially reduces the unphysical artifacts relative to plain superposition. As shown in Ref.~\cite{Atteneder:2023pge}, the dominant ingredient for improving the physical diagnostics, including the scalar-field dynamics and the extracted GW signal, is the volume-element correction at the BS center, while subsequent elliptic solving of the constraint equations provides only a subdominant improvement. In addition, the CCZ4 formulation used in our simulations includes built-in constraint damping, which further mitigates the impact of the residual initial violations on the subsequent evolution. We therefore leave the construction of fully constraint-solved initial data for future work and expect the improved superposition method to provide a substantial improvement over plain superposition in our BS-BH binary simulations.

\section{Numerical results} \label{sec:results}

Before presenting the results of our BS-BH binary simulations, we first summarize the initial-data configurations considered in this work. We focus on solitonic BSs with $\sigma_0 = 0.2$, which allows us to study relatively compact configurations. We consider three BS models with central amplitudes $A(0) = 0.17$, $0.147$, and $0.1$, corresponding to highly compact, moderately compact, and low-compactness BSs, respectively. For simplicity, we label them H, M, and L. The corresponding properties of these BSs are listed in Table~\ref{BSs}.

\begin{table}[htbp]
    \centering
    \begin{ruledtabular}
        \begin{tabular}{ccccccc}
            Model & $A(0)$ & $\mu M_\mathrm{BS}$ & $\omega/\mu$ & $\mu r_{99}$ & $\mathcal{C}$ \\
            \hline
            \texttt{BS-170} & 0.17 & 0.7134 & 0.4392 & 3.3224 & 0.2146\\
            \texttt{BS-147} & 0.147 & 0.3609 & 0.6780 & 4.2557 & 0.0848 \\
            \texttt{BS-100} & 0.1 & 0.2700 & 0.8506 & 6.1717 & 0.0437 \\
        \end{tabular}
    \end{ruledtabular}
    \caption{Solitonic BS models with $\sigma_0=0.2$ considered in this work. Here, $A(0)$ denotes the central scalar-field amplitude, $M_\mathrm{BS}$ the ADM mass of the BS, $\omega$ the frequency of the ground-state solution, $r_{99}$ the radius (in isotropic coordinates) containing $99\%$ of the BS mass, and $\mathcal{C} \equiv M_\mathrm{BS}/r_{99}$ is a measure of compactness. For this potential, the maximum BS mass is $\mu M_\mathrm{BS} = 0.7212$.}
    \label{BSs}
\end{table}

For the BS-BH binaries, we consider mass ratios $q = M_\mathrm{BS}/M_\mathrm{BH} = 0.38$, $0.5$, $1$, $2$, and $2.64$, and we also include several BS-BS and BH-BH cases for comparison. To test the validity of the improved superposition method, we additionally include a plain-superposition case for the equal-mass BS-BH binary and vary the initial separation to assess the severity of the artifacts. The parameters of the binary configurations are listed in Table~\ref{binarys}. The initial velocities are chosen so that the total initial momentum vanishes, while the initial relative velocity is kept fixed at $\Delta v = v_A - v_B = 0.2$ for all cases. For all BS-BH binaries, we use the improved superposition method described in Sec.~\ref{sec:improved_superposition}. For \texttt{BSBH-q1-H}, we also include the plain superposition method to illustrate the severity of the artifacts it produces. For the BS-BS binaries, we use the improved superposition method of Ref.~\cite{Evstafyeva:2022bpr} for all cases. We uniformly set the correction-radius parameter to $R_\mathrm{fix} = 50$ for all BSs in the fiducial sample. We have also varied the correction radius to $R_\mathrm{fix} = 25$ and $100$ for the $q=0.5$ and $q=2.64$ cases to assess the sensitivity of the results to this parameter. Across the corrected profiles, the total GW energy varies at the percent level, while the principal difference in the dominant $(2,0)$ waveform is a timing shift. For the BH-BH binaries, we use the plain-superposition construction as a reference baseline; for the separations and velocities considered here, this provides a sufficient comparison dataset for the purposes of this work. For the reciprocal $q = 0.38$ and $q = 2.64$ comparison, all runs are evolved with a base grid of $N = 512$, twice the fiducial global resolution, and the runs containing the $M_\mathrm{BH} = 0.2700$ puncture use one additional AMR level.

\begin{table*}[htbp]
    \centering
    \begin{ruledtabular}
        \begin{tabular}{ccccccccc}
            Run & $q$ & Object A & Object B & $v_{x}^\mathrm{A}$ & $v_{x}^\mathrm{B}$ & $d$ &$M$ & Initial data \\
            \hline
            \texttt{BSBH-q1-H} & $1$ & \texttt{BS-170} & $M_\mathrm{BH} = 0.7134$ & $0.1$ & $-0.1$ & $20, 40, 60, 80$ & $1.4340$ & Improved/Plain \\
            \texttt{BSBH-q1-M} & $1$ & \texttt{BS-147} & $M_\mathrm{BH} = 0.3609$ & $0.1$ & $-0.1$ & $40$ & $0.7254$ & Improved \\
            \texttt{BSBH-q1-L} & $1$ & \texttt{BS-100} & $M_\mathrm{BH} = 0.2700$ & $0.1$ & $-0.1$ & $40$ & $0.5427$ & Improved \\
            \texttt{BSBH-q038} & $0.38$ & \texttt{BS-100} & $M_\mathrm{BH} = 0.7134$ & $0.1447$ & $-0.0553$ & $40$ & $0.9874$ & Improved \\
            \texttt{BSBH-q05} & $0.5$ & \texttt{BS-147} & $M_\mathrm{BH} = 0.7134$ & $0.1325$ & $-0.0675$ & $40$ & $1.0791$ & Improved \\
            \texttt{BSBH-q2} & $2$ & \texttt{BS-170} & $M_\mathrm{BH} = 0.3609$ & $0.0675$ & $-0.1325$ & $40$ & $1.0791$ & Improved \\
            \texttt{BSBH-q2.64} & $2.64$ & \texttt{BS-170} & $M_\mathrm{BH} = 0.2700$ & $0.0553$ & $-0.1447$ & $40$ & $0.9874$ & Improved \\

            \texttt{BSBS-q1-H} & $1$ & \texttt{BS-170} & \texttt{BS-170} & $0.1$ & $-0.1$ & $40$ & $1.4340$ & Improved \\
            \texttt{BSBS-q1-M} & $1$ & \texttt{BS-147} & \texttt{BS-147} & $0.1$ & $-0.1$ & $40$ & $0.7254$ & Improved \\
            \texttt{BSBS-q1-L} & $1$ & \texttt{BS-100} & \texttt{BS-100} & $0.1$ & $-0.1$ & $40$ & $0.5427$ & Improved \\
            \texttt{BSBS-q038} & $0.38$ & \texttt{BS-100} & \texttt{BS-170} & $0.1447$ & $-0.0553$ & $40$ & $0.9874$ & Improved \\
            \texttt{BSBS-q05} & $0.5$ & \texttt{BS-147} & \texttt{BS-170} & $0.1325$ & $-0.0675$ & $40$ & $1.0791$ & Improved \\

            \texttt{BHBH-q1} & $1$ & $M_\mathrm{BH} = 0.7134$ & $M_\mathrm{BH} = 0.7134$ & $0.1$ & $-0.1$ & $40$ & $1.4340$ & Plain \\
            \texttt{BHBH-q038} & $0.38$ & $M_\mathrm{BH} = 0.2700$ & $M_\mathrm{BH} = 0.7134$ & $0.1447$ & $-0.0553$ & $40$ & $0.9874$ & Plain \\
            \texttt{BHBH-q05} & $0.5$ & $M_\mathrm{BH} = 0.3609$ & $M_\mathrm{BH} = 0.7134$ & $0.1325$ & $-0.0675$ & $40$ & $1.0791$ & Plain \\
        \end{tabular}
    \end{ruledtabular}
    \caption{Binary initial-data configurations considered in this work. Each run label begins with the binary type and ends with a suffix denoting the mass ratio $q = M_\mathrm{A}/M_\mathrm{B}$. Here, $v_x^\mathrm{J}$ for $\mathrm{J} \in \{\mathrm{A}, \mathrm{B}\}$ denotes the initial boost velocity, with Lorentz factor $\gamma_\mathrm{J} = 1/\sqrt{1 - {v_x^\mathrm{J}}^2}$. The initial velocities are chosen so that the total initial momentum $P_x = M_\mathrm{A} \gamma_\mathrm{A} v_x^\mathrm{A} + M_\mathrm{B} \gamma_\mathrm{B} v_x^\mathrm{B}$ vanishes, while the initial relative velocity is fixed at $\Delta v = v_x^\mathrm{A} - v_x^\mathrm{B} = 0.2$ for all runs. The total mass of the binary is denoted by $M = \gamma_\mathrm{A} M_\mathrm{A} + \gamma_\mathrm{B} M_\mathrm{B}$. For the BS-BH binaries, we use the improved superposition method described in Sec.~\ref{sec:improved_superposition} for all runs except \texttt{BSBH-q1-H}, for which we also include the plain superposition method to demonstrate its artifacts. For the BS-BS binaries, we use the improved superposition method of Ref.~\cite{Evstafyeva:2022bpr} for all runs. For the BH-BH binaries, we use the plain superposition method for all runs, since it does not suffer from the BS-specific superposition artifacts.}
    \label{binarys}
\end{table*}

\subsection{Comparison of initial data}

\subsubsection{Initial constraint violations}

\begin{figure*}[htbp]
    \centering
    \includegraphics[width=0.45\linewidth]{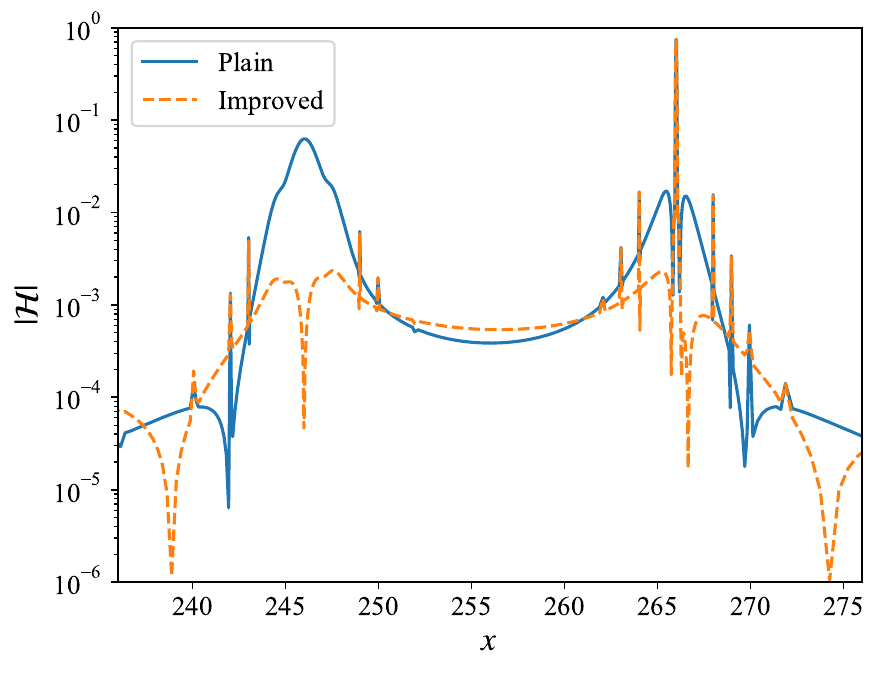}
    \includegraphics[width=0.45\linewidth]{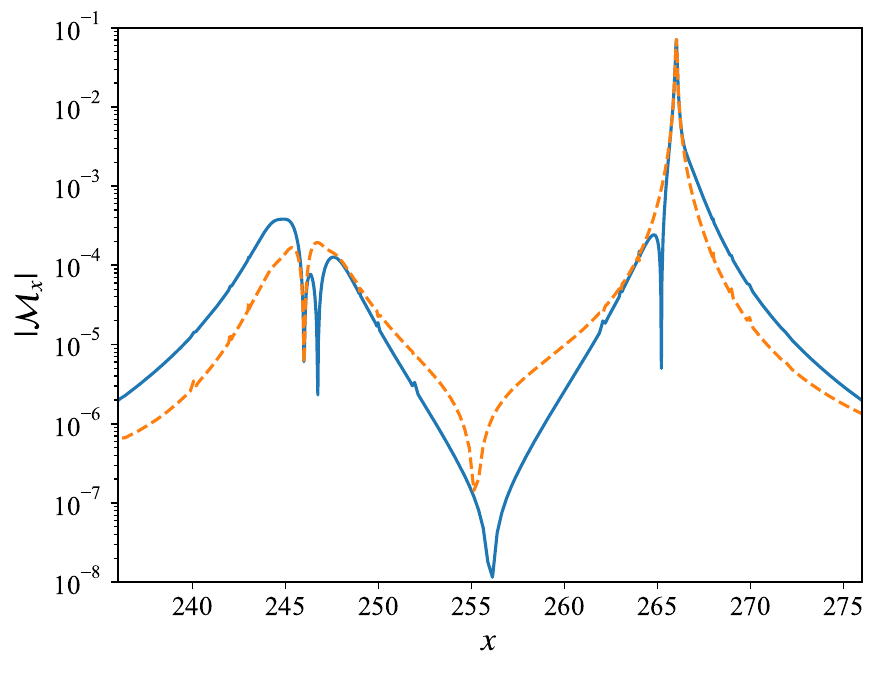}
    \caption{Comparison of the initial Hamiltonian (left panel) and momentum (right panel) constraint violations along the collision axis for the configuration \texttt{BSBH-q1-H} with $d = 20$, generated using the plain (solid lines) and improved (dashed lines) superposition methods, respectively. The improved superposition method significantly reduces the Hamiltonian-constraint violation in the vicinity of both the BS (at $x = 246$) and the BH (at $x = 266$) relative to the plain superposition method.}
    \label{fig:initial_constraint}
\end{figure*}

Since the plain superposition method can distort the volume element at the BS center and thereby inject spurious mass-energy, one expects the improved superposition method to reduce the Hamiltonian-constraint violation significantly. To demonstrate this, we plot the initial Hamiltonian-constraint violation along the collision axis for the configuration \texttt{BSBH-q1-H} with $d = 20$ in the top panel of Fig.~\ref{fig:initial_constraint}, comparing the plain and improved superposition methods with solid and dashed lines, respectively. We find that the improved superposition method reduces the Hamiltonian-constraint violation at the BS center (at $x = 246$) by several orders of magnitude relative to the plain superposition method. Moreover, the Hamiltonian-constraint violation is also substantially reduced in the vicinity of the BH puncture (at $x = 266$), even though no correction is applied there (unlike the \texttt{TwoPunctures} correction used in Ref.~\cite{Marks:2026xvo}). The momentum-constraint violation, shown in the bottom panel of Fig.~\ref{fig:initial_constraint}, is also mildly reduced near the BS. These results demonstrate the effectiveness of the improved superposition method in reducing strong-field constraint violations and preserving the BS center volume element. However, the correction has a long spatial tail and can increase the much smaller Hamiltonian residual in the wave zone. Appendix~\ref{app:convergence} therefore separately reports the constraint violations there, which remain at the few-$10^{-6}$ level during the evolution.

\subsubsection{Evolution of the scalar field and premature collapse}

\begin{figure}[htbp]
    \centering
    \includegraphics[width=0.9\linewidth]{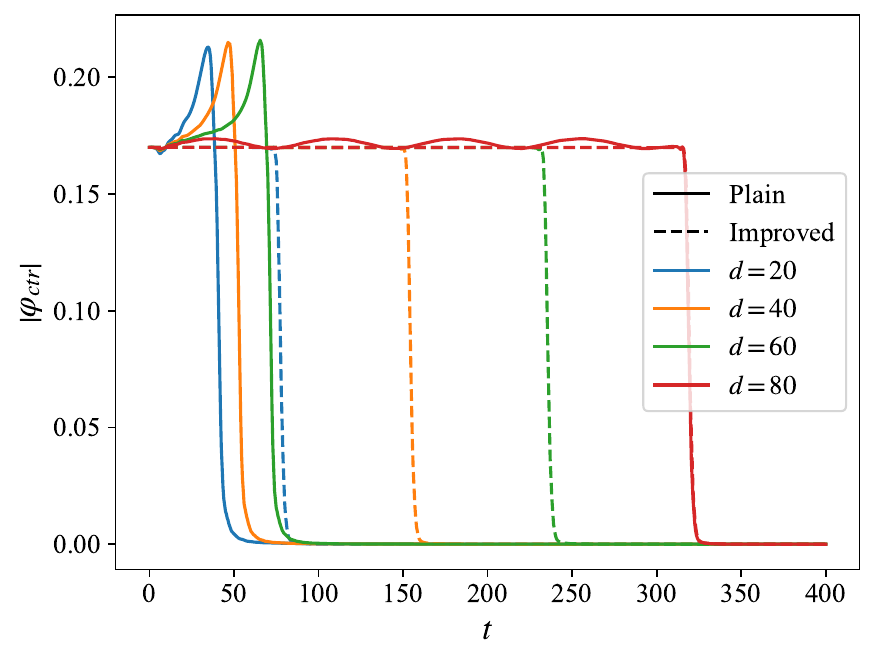}
    \caption{Time evolution of the central scalar-field amplitude $|\varphi_c(t)|$ for the configuration \texttt{BSBH-q1-H}, generated using the plain (solid lines) and improved (dashed lines) superposition methods at different initial separations. The plain superposition method leads to strong unphysical radial oscillations and, for smaller initial separations, premature collapse of the BS into a BH. The improved superposition method removes these artifacts and keeps the scalar amplitude close to its initial equilibrium value until the physical merger dynamics begins.}
    \label{fig:mod_phi_ctr}
\end{figure}

We now compare the time evolution of the scalar field for the configuration \texttt{BSBH-q1-H} generated using the plain and improved superposition methods. The model \texttt{BS-170} is a highly compact BS, lying below but close to the maximum-mass threshold and therefore close to the instability threshold. Fig.~\ref{fig:mod_phi_ctr} shows the time evolution of the central scalar-field amplitude $|\varphi_c(t)|$ for the plain (solid lines) and improved (dashed lines) superposition methods at different initial separations. For the plain-superposition runs with $d = 20$, $40$, and $60$, the scalar amplitude increases significantly from its initial value and then rapidly drops to a near-zero level, indicating premature collapse of the BS into a BH before the physical merger dynamics begins. The formation of a BH is also confirmed by the horizon finder. For the largest initial separation, $d = 80$, premature collapse is avoided, but the scalar amplitude still exhibits significant unphysical radial oscillations, with the equilibrium value acting as the minimum value. In all cases, the improved superposition method removes the unphysical oscillations and prevents premature collapse, keeping the scalar amplitude close to its initial equilibrium value until the physical merger dynamics begins. These results demonstrate that BS-BH collisions exhibit the same type of spurious behavior caused by plain superposition as in BS-BS collisions, and that the improved superposition method effectively mitigates these artifacts.

\subsubsection{Radiated gravitational waveforms and energy}

\begin{figure}[htbp]
    \centering
    \includegraphics[width=0.9\linewidth]{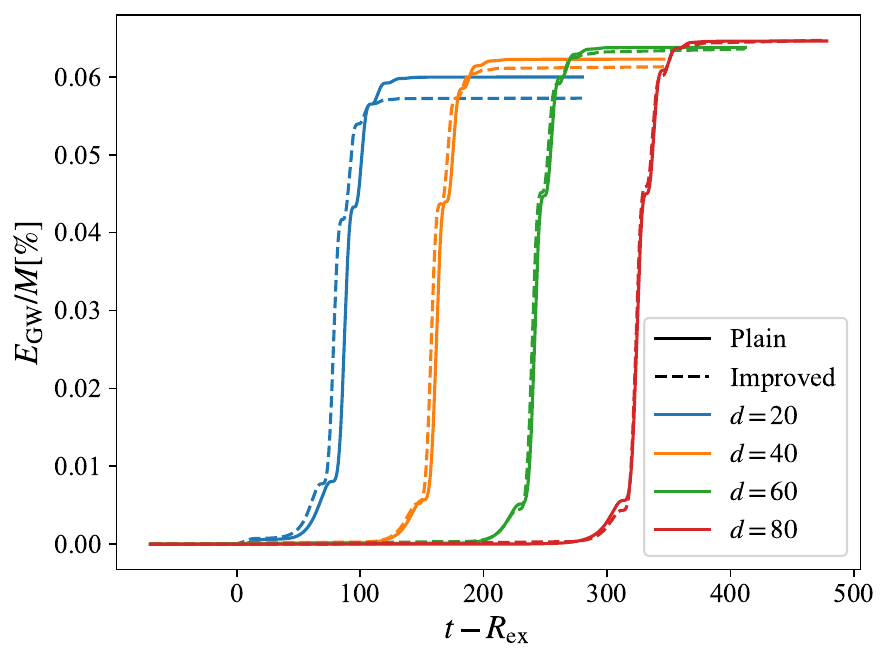}
    \\
    \includegraphics[width=0.9\linewidth]{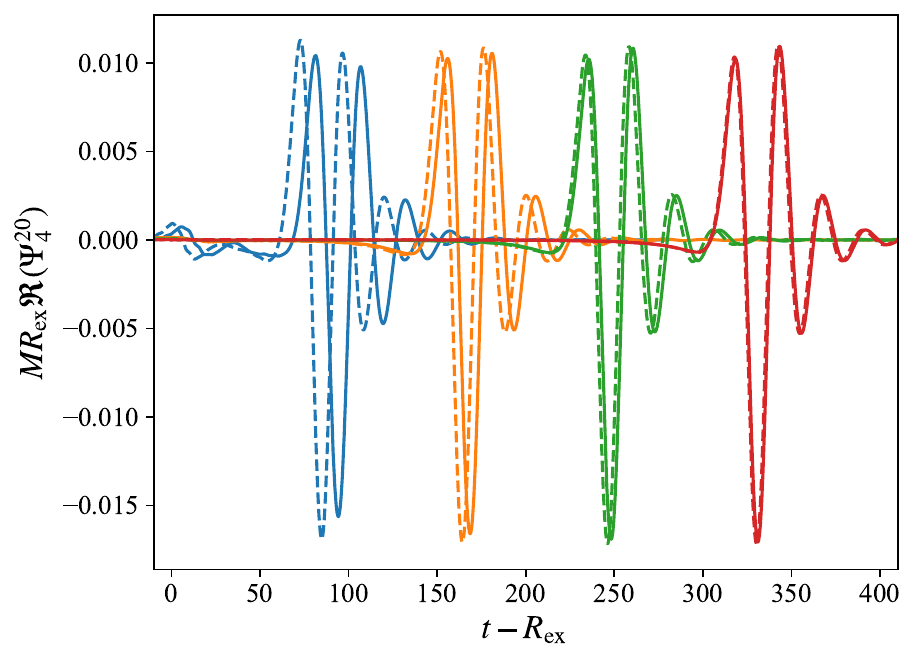}
    \caption{Comparison of the radiated GW energy (top panel) and the $(2,0)$ mode of the Newman-Penrose scalar $\Psi_4$ (bottom panel) for the configuration \texttt{BSBH-q1-H}, generated using the plain (solid lines) and improved (dashed lines) superposition methods at different initial separations. Relative to the improved method, the plain superposition method overestimates the radiated energy and introduces unphysical distortions in the waveform, especially at smaller initial separations. As the initial separation increases, the two methods yield more compatible results.}
    \label{fig:q1-dX_GW}
\end{figure}

We now present the GW waveforms and radiated energy for the configuration \texttt{BSBH-q1-H}, comparing the plain and improved superposition methods at different initial separations. Because the plain superposition method injects spurious mass-energy into the BS and even triggers premature collapse at smaller separations, it is expected to produce more GW energy than the improved superposition method, which provides more accurate initial data. This is confirmed by the results shown in the top panel of Fig.~\ref{fig:q1-dX_GW}. The increase in the radiated energy with increasing initial separation is physical and is associated with the larger collision velocity at merger due to the longer free-fall time, as discussed in Ref.~\cite{Evstafyeva:2022bpr} for the BS-BS case. However, the plain superposition method overestimates the radiated energy relative to the improved method, especially for smaller initial separations. As the initial separation increases, the difference between the two methods becomes less pronounced, and for $d = 80$ both approaches yield compatible results because the BH gravitational tail has sufficiently decayed at the BS center.

In the bottom panel of Fig.~\ref{fig:q1-dX_GW}, we show the $(2,0)$ mode of the Newman-Penrose scalar $\Psi_4$ for the same configurations. The amplitude of the $(2,0)$ mode is slightly higher for the improved superposition method than for the plain superposition method, although the difference is not large. However, the waveforms produced by the plain superposition method are significantly stretched in time relative to those from the improved method, which leads to a larger radiated energy after time integration. This stretching effect is more pronounced at smaller initial separations, whereas for larger separations the waveforms from the two methods become more similar. These results show that the plain superposition method not only overestimates the radiated energy but also introduces unphysical distortions into the waveform. The improved superposition method is therefore essential for obtaining accurate GW signals and energy estimates from BS-BH binary simulations, particularly when the initial separation is small and the superposition artifacts are strongest.

\subsection{Equal-mass BS-BH collisions}

After demonstrating the effectiveness of the improved superposition method in mitigating artifacts and obtaining more accurate GWs for BS-BH binaries, we now present the results of our BS-BH simulations. We first focus on the equal-mass case with $q = 1$ and compare the gravitational waveforms for BS-BH, BS-BS, and BH-BH collisions across different BS models.

\begin{figure}[htbp]
    \centering
    \includegraphics[width=0.9\linewidth]{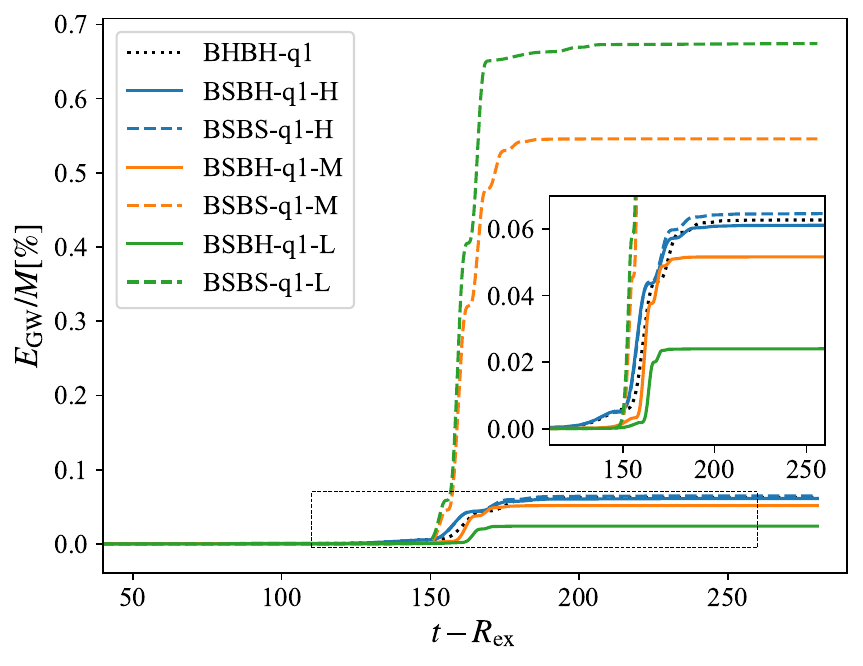}
    \\
    \includegraphics[width=0.9\linewidth]{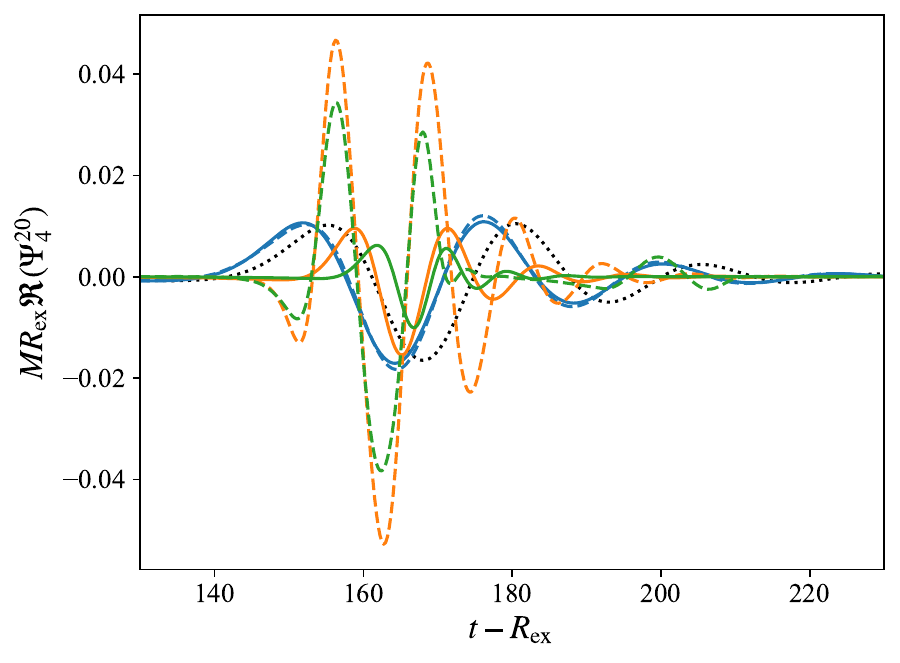}
    \caption{Comparison of the radiated GW energy (top panel) and the dominant $(2,0)$ mode of $\Psi_4$ (bottom panel) for equal-mass BS-BH, BS-BS, and BH-BH collisions with different BS models.}
    \label{fig:equal}
\end{figure}

The top panel of Fig.~\ref{fig:equal} shows the radiated GW energy for equal-mass BS-BH, BS-BS, and BH-BH collisions with an initial separation of $d = 40$. For the BS-BH collisions (solid lines), our results agree with the recent findings of Marks et al.~\cite{Marks:2026xvo}: the GW radiation efficiency increases with the compactness of the BS, and the most compact BS-BH system approaches the BH-BH limit from below (see the \texttt{BS-170} model). By contrast, for the BS-BS collisions (dashed lines) with the same BS models, a strikingly different trend emerges: in all three runs, the radiated energy exceeds that of the BH-BH case and decreases as the compactness increases. Moreover, the radiated energy for the medium- and low-compactness BS-BS collisions is much larger than that of the BH-BH case. This behavior reflects the fact that GW emission efficiency is determined not only by compactness, but also by the tidal deformability~\cite{Sennett:2017etc} of the compact objects. For a detailed discussion of GW emission from BS-BS binaries, we refer the reader to Refs.~\cite{Ge:2024fum,Ge:2024itl,Ge:2025btw}. In contrast, in BS-BH collisions the BH rapidly accretes the extended scalar cloud of the BS, thereby suppressing the kinetic energy that would otherwise be radiated as GWs. Therefore, one should not naively expect the GWs emitted by a BS-BH binary to lie between those emitted by the corresponding BS-BS and BH-BH systems.

The bottom panel of Fig.~\ref{fig:equal} shows the $(2,0)$ mode of the Newman-Penrose scalar $\Psi_4$ for the same configurations. All three binary types exhibit a similar qualitative structure---a primary merger pulse followed by a ringdown phase---but they differ markedly in their peak amplitudes and pulse widths. For the highly compact case (\texttt{H}), the BS-BH and BS-BS waveforms are morphologically almost indistinguishable from the BH-BH baseline, reinforcing the ``BH mimicker'' paradigm in the highly compact regime. However, for the less compact configurations (\texttt{M} and \texttt{L}), a pronounced contrast appears. The BS-BS collisions exhibit a strongly amplified and sharply peaked primary pulse, whereas the amplitudes of the corresponding BS-BH waveforms are substantially suppressed.

\begin{figure}[htbp]
    \centering
    \includegraphics[width=0.9\linewidth]{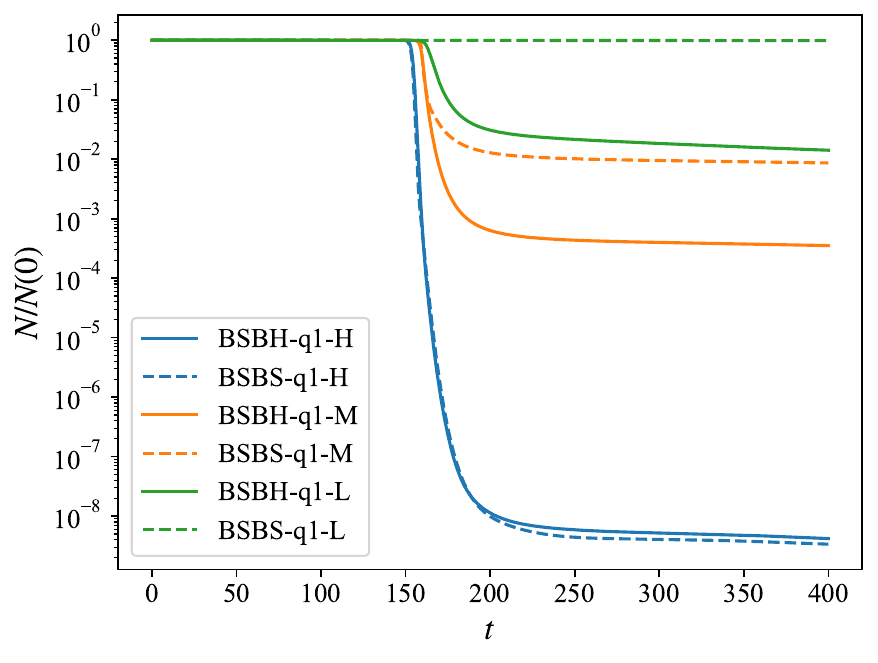}
    \caption{Time evolution of the Noether charge $N$ for equal-mass BS-BH (solid lines) and BS-BS (dashed lines) collisions with different BS models. For the BS-BH collisions, the residual grid-integrated Noether charge after the merger increases as the compactness of the BS decreases.}
    \label{fig:NoetherCharge}
\end{figure}

We also plot the time evolution of the total grid-integrated Noether charge $N$ for the equal-mass BS-BH and BS-BS collisions in Fig.~\ref{fig:NoetherCharge}. For the BS-BH collisions, the Noether charge nearly vanishes after the merger, consistent with substantial scalar-field accretion. However, this global diagnostic alone does not separate horizon flux, outer-boundary flux, and numerical loss, and we therefore do not infer an accreted fraction or an exterior charge fraction from Fig.~\ref{fig:NoetherCharge}. A finite-volume balance for representative unequal-mass BS-BH configurations is presented in Appendix~\ref{app:matter_diagnostics}. Moreover, the residual Noether charge increases as the compactness of the BS decreases, consistent with Refs.~\cite{Clough:2018exo,Marks:2026xvo}. For the BS-BS collisions, the Noether charge also nearly vanishes after the merger when the final state is a BH (for the high- and medium-compactness cases), whereas for the low-compactness case the final state is a BS, and the Noether charge remains constant after the merger.

\subsection{Unequal-mass BS-BH collisions}

\begin{figure*}[htbp]
    \centering
    \includegraphics[width=0.45\linewidth]{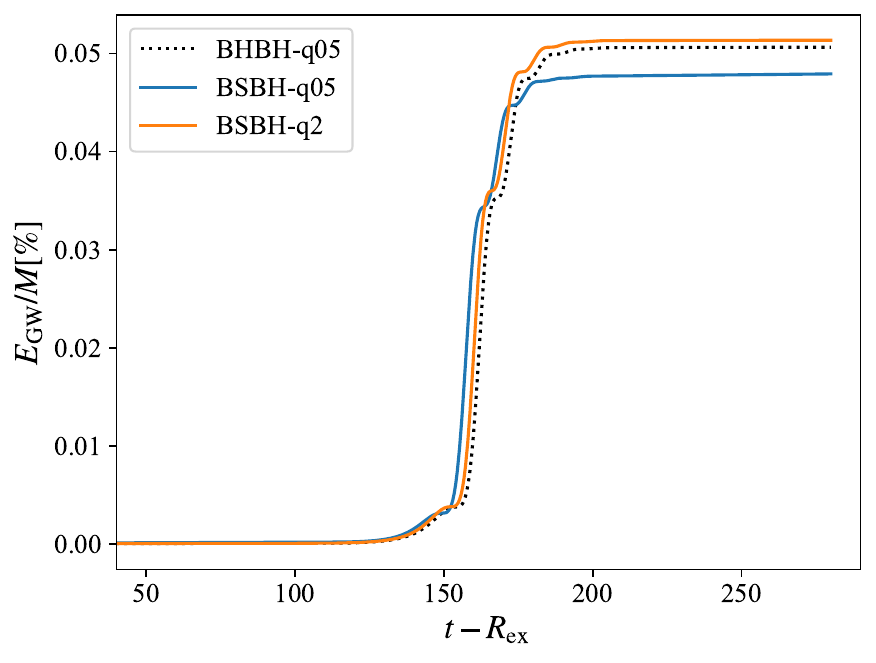}
    \includegraphics[width=0.45\linewidth]{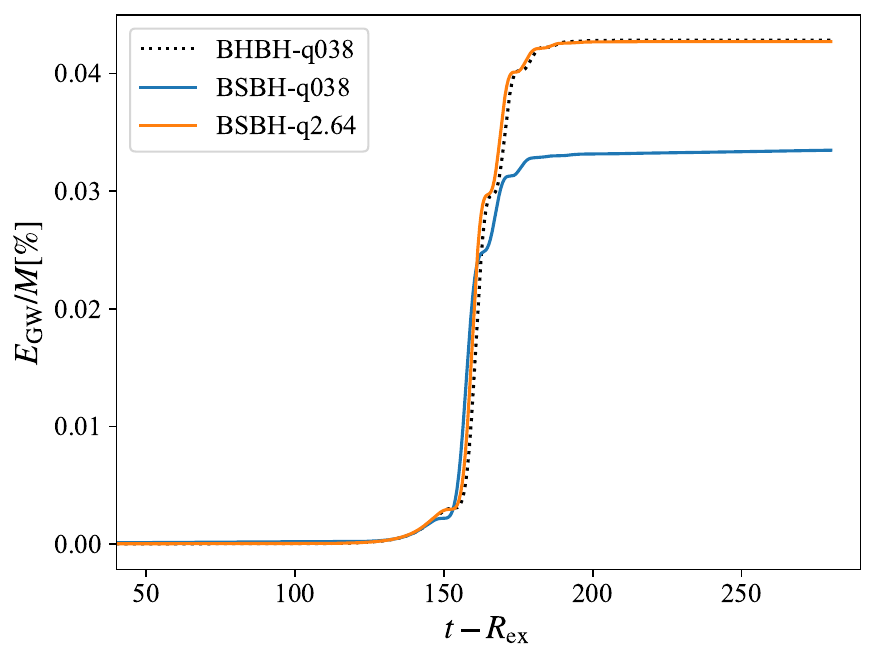}
    \\
    \includegraphics[width=0.45\linewidth]{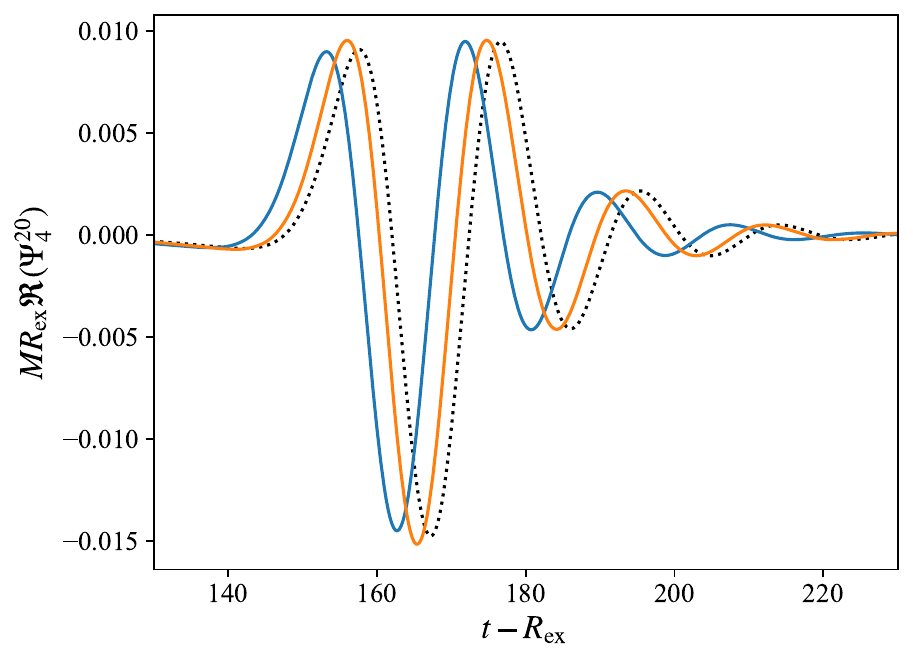}
    \includegraphics[width=0.45\linewidth]{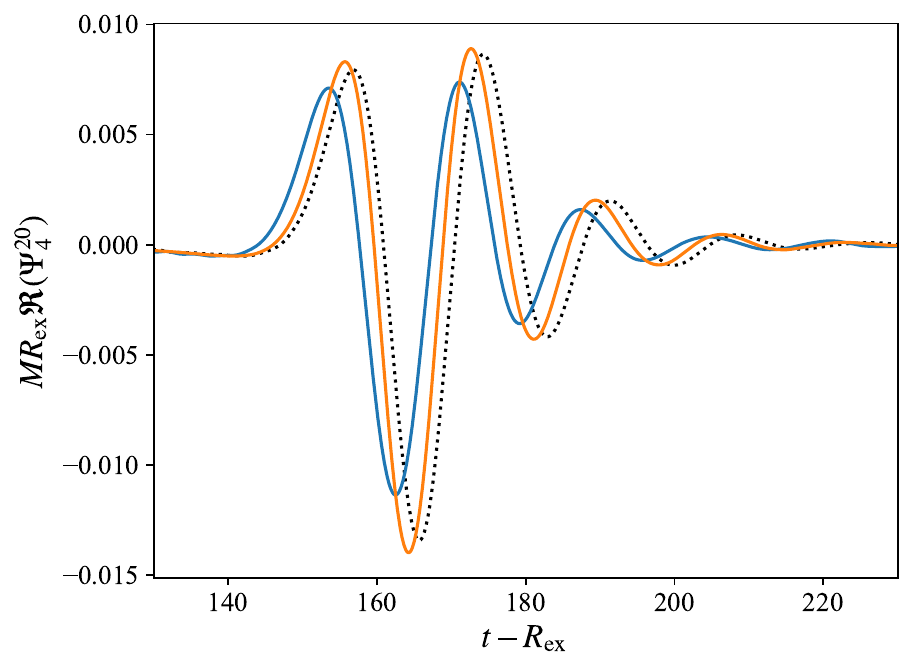}
    \caption{Comparison of the radiated GW energy (top panels) and the dominant $(2,0)$ mode of $\Psi_4$ (bottom panels) for unequal-mass BS-BH (solid lines) and BH-BH (dotted lines) collisions with mass ratios $q \in \{0.5, 2\}$ (left column) and $q \in \{0.38, 2.64\}$ (right column).}
    \label{fig:unequal}
\end{figure*}

Before examining the GW emission from unequal-mass binaries, we must address a major distinction between BS-BS and BS-BH systems in their initial parameter space. As emphasized by Evstafyeva et al.~\cite{Evstafyeva:2022bpr}, the merger dynamics and radiated energy of unequal-mass BS-BS binaries are highly sensitive to the relative scalar phase at the moment of collision. Because unequal-mass BSs have different characteristic frequencies ($\omega_A \neq \omega_B$), they inevitably accumulate a dynamical phase difference $\Delta\phi \approx (\omega_A - \omega_B)t$ during the infall. Consequently, setting the initial phase difference to zero at $t = 0$ leads to a merger phase---and therefore to constructive or destructive interference patterns---that depends strongly on the initial separation $d$. This strong phase dependency makes it difficult to define a representative GW energy for unequal-mass BS-BS collisions without an exhaustive scan over the phase space.

By contrast, the BH in a BS-BH binary carries no scalar phase. The interaction and accretion dynamics are governed solely by the BS stress-energy tensor, which is invariant under global $U(1)$ phase transformations. The merger dynamics of BS-BH systems are therefore free from any scalar-phase dependency. To ensure a rigorous and unambiguous comparison across binary parameters, we exclude the strongly phase-dependent BS-BS configurations from the following analysis of unequal-mass systems. Instead, we focus on contrasting the phase-independent BS-BH collisions with their pure BH-BH counterparts across different mass ratios $q$.

We first examine the radiated GW energy, shown in the top panels of Fig.~\ref{fig:unequal} for mass ratios $q \in \{0.5, 2\}$ (left) and $q \in \{0.38, 2.64\}$ (right). When the BH is the heavier companion ($q < 1$, blue solid lines), the radiative efficiency of the BS-BH binary is consistently lower than that of the corresponding BH-BH system (black dotted lines), as in the equal-mass case. However, a more intricate nonlinear competition emerges when the highly compact BS is the heavier object ($q > 1$, orange solid lines). For $q = 2$, the radiated energy of the BS-BH system slightly exceeds that of the pure BH-BH collision. This is consistent with the recent findings of Marks et al.~\cite{Marks:2026xvo}, who observed that sufficiently compact BS-BH collisions can radiate more efficiently than their BH-BH counterparts in the $q > 1$ regime. Interestingly, when the mass ratio is increased further to $q = 2.64$, the BS-BH GW energy is comparable to the matched BH-BH baseline, rather than showing a further enhancement. The GW energy ratio $E_{\rm GW}^{q=2.64}/E_{\rm GW}^{\rm BHBH}$ is $0.98225$ at the fiducial resolution and is $0.99725$ on the twice-finer global grid, so still higher resolution would be required to resolve such a small difference conclusively. This behavior appears to contrast with Ref.~\cite{Marks:2026xvo}, where the positive excess in radiated energy for $q > 1$ was found to grow with $q$ up to $q = 3$. This difference may be attributed to the distinct kinematic setups or the different solitonic BS models and compactnesses. Marks et al. adopted a constant initial velocity ($v = 0.1$) for all objects, while we impose a center-of-momentum configuration with $P_x = 0$ and $\Delta v = 0.2$.

Despite these differences in the total radiated energy, the dominant $(2,0)$ quadrupolar waveforms (bottom panels of Fig.~\ref{fig:unequal}) remain morphologically very similar across all configurations. Their temporal width, merger peak, and ringdown frequencies closely track those of the pure BH-BH baseline. This reaffirms the paradigm that, in the dominant GW channel, highly compact BSs are excellent BH mimickers, even in the unequal-mass regime.

\begin{figure*}[htbp]
    \centering
    \includegraphics[width=\linewidth]{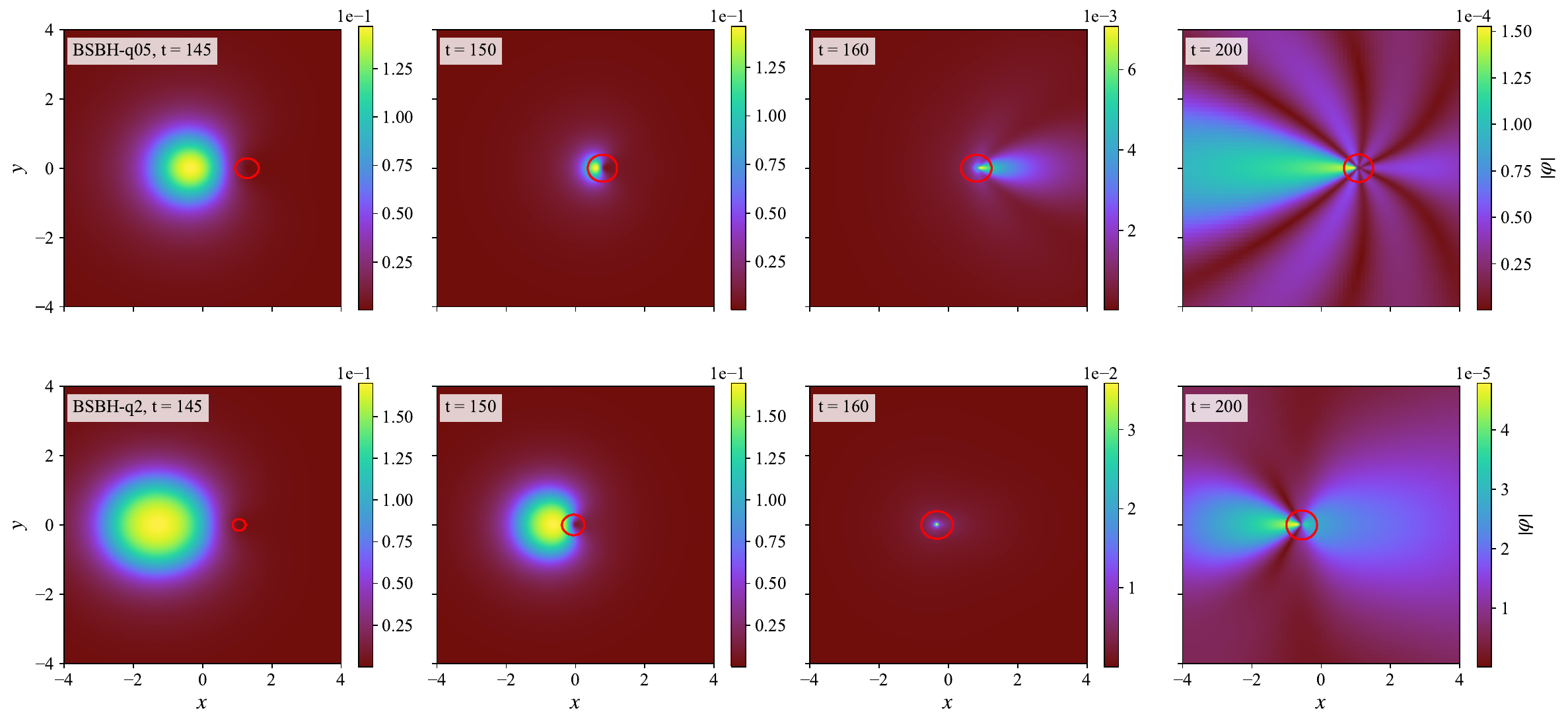}
    \caption{Cross-sectional snapshots of the scalar-field modulus $|\varphi|$ in the $x$-$y$ plane for unequal-mass BS-BH collisions with mass ratios $q = 0.5$ (top row) and $q = 2$ (bottom row) at different times. The red circle marks the contour of constant conformal factor $\chi = 1/16$, which corresponds to the value at the event horizon for a Schwarzschild BH in isotropic coordinates and serves as a rough estimate of the horizon location. The snapshots are shown only as qualitative illustrations of the scalar-accretion evolution.}
    \label{fig:slice}
\end{figure*}

The scalar-field modulus cross-sections in Fig.~\ref{fig:slice} provide qualitative illustrations of scalar accretion in the representative $q = 0.5$ and $q = 2$ collisions. The successive panels show the approach to contact, compression of the BS, and transfer of the scalar field into the BH region as each collision proceeds. The integrated exterior/cloud Noether charges associated with the late-time residual scalar field at $t = 200$ lie below the late-time charge-conservation floor derived in Appendix~\ref{app:matter_diagnostics}; we therefore do not interpret them as resolved long-lived gravitational atoms.

\section{Conclusions and discussion} \label{sec:conclusions}

In this work, we have presented a systematic numerical-relativity study of comparable-mass BS-BH head-on collisions, with emphasis on both initial-data construction and GW phenomenology. Our main technical advance is an improved BS-centered one-body conformal-factor correction for BS-BH binaries, obtained as the BS-BH limit of the unequal-mass BS-BS prescription. This construction is designed to restore the BS core volume element while preserving the plain-superposition structure away from the star.

Our principal findings can be summarized as follows.

First, at the level of the initial data, the improved prescription significantly suppresses the Hamiltonian-constraint violation near the BS core, and it also reduces the violation around the BH puncture, compared with plain superposition. More importantly, it removes the spurious early collapse and large unphysical oscillations of the central scalar amplitude that appear in plain-superposition evolutions for compact stars at modest separations.

Second, for equal-mass binaries, BS-BH collisions show a clear compactness trend: the GW radiation efficiency increases with BS compactness and approaches the BH-BH limit for highly compact stars. In contrast, the equal-mass BS-BS collisions in our sample can radiate more strongly than the BH-BH case and follow a different compactness trend, highlighting that compactness alone does not control the emission efficiency across different binary classes.

Third, for unequal-mass BS-BH binaries, the dominant $(2,0)$ mode remains close to the BH-BH morphology in many cases. Within our unequal-mass sample, the $q = 0.5$ and $q = 0.38$ BS-BH configurations radiate less GW energy than their matched BH-BH baselines, the $q = 2$ configuration radiates slightly more, and the $q = 2.64$ configuration is comparable at the highest available resolution.

These results show that high compactness can make BS-BH head-on collisions partially degenerate with BH-BH systems in their total radiated energy and dominant quadrupolar emission. However, our study also has clear limitations. We have focused on head-on collisions, a single scalar-potential family, and a finite set of mass ratios and compactnesses. The axisymmetry and reflection properties of a head-on collision differ from those of a quasi-circular inspiral, so our results should not be extrapolated to detector-level distinguishability. Extending this program to quasi-circular inspirals, broader potential models, including mini-BS baselines, and a larger parameter-space coverage is a natural next step. It will also be important to quantitatively compare against fully constraint-solved initial data and different initial-data correction prescriptions, such as the method of Ref.~\cite{Marks:2026xvo}.

\noindent {\bf \em Note added.}
While we were finalizing this work, Ref.~\cite{Marks:2026xvo} appeared on arXiv, which also studied comparable-mass BS-BH binaries using numerical-relativity simulations. Their study and ours overlap in the broad goal of characterizing GW emission from BS-BH mergers, but they are complementary in methodology and diagnostics. First, we employ a BS-centered one-body conformal-factor correction derived from the unequal-mass BS-BS prescription, whereas Ref.~\cite{Marks:2026xvo} uses a different initial-data treatment that combines a BS metric correction with a BH \texttt{TwoPunctures} correction. Second, we cross-compare the GW emission from BS-BH binaries with both BS-BS and BH-BH counterparts at matched kinematics, whereas Ref.~\cite{Marks:2026xvo} focuses on the comparison between BS-BH and BH-BH binaries. In the overlapping regimes, both studies indicate that compactness is a key control parameter for GW emission, and that highly compact BS-BH systems can approach BH-BH-like GW signatures. Quantitative differences in some unequal-mass cases are plausibly attributable to different initial-data prescriptions, initial kinematic setups, and solitonic BS models and compactnesses. Overall, both studies provide valuable insights into GW emission from BS-BH binaries and highlight the importance of accurate initial data and comprehensive diagnostics for understanding these complex systems.

\begin{acknowledgments}
    This work is supported in part by the National Key Research and Development Program of China Grant No. 2021YFC2203004, No. 2021YFA0718304, and No. 2020YFC2201501, the National Natural Science Foundation of China Grants No. 12422502, No. 12547110, No. 12588101, No. 12235019, and No. 12447101, and the Science Research Grants from the China Manned Space Project with No. CMS-CSST-2025-A01.
\end{acknowledgments}

\section*{Data Availability}

The data that support the findings of this article are openly available~\cite{Ning_BSBH_Data_2026}.

\appendix

\section{Convergence and wave-zone constraint tests} \label{app:convergence}

\begin{figure}[htbp]
    \centering
    \includegraphics[width=0.9\linewidth]{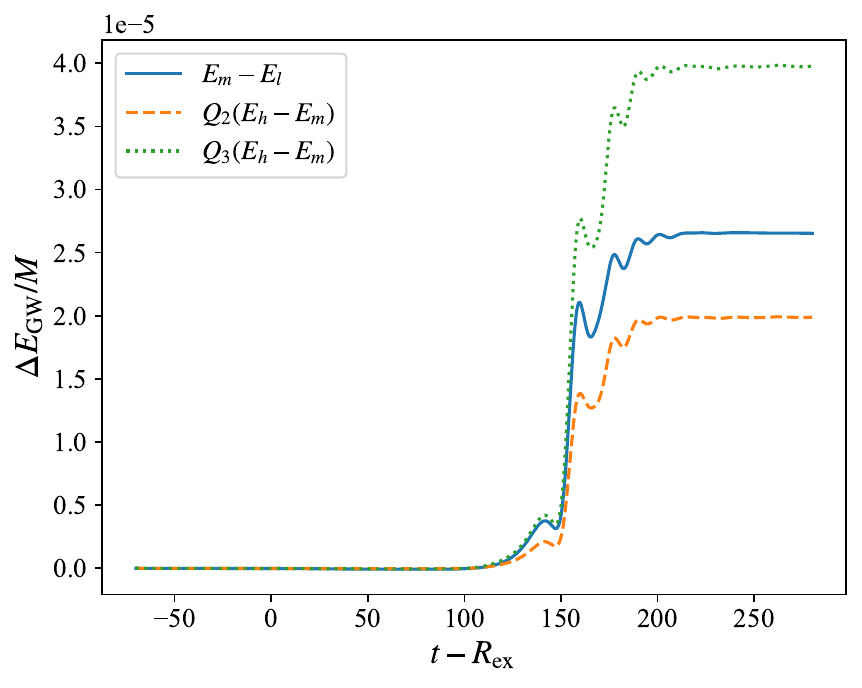}
    \caption{Convergence test of the radiated GW energy for the run \texttt{BSBH-q1-H} with $d = 40$ at three different resolutions. The differences between the high- and medium-resolution results are multiplied by the convergence factor $Q_p$ for $p = 2$ and $p = 3$. The convergence order is estimated to lie between 2 and 3.}
    \label{fig:convergence_GW}
\end{figure}

\begin{figure*}[htbp]
    \centering
    \includegraphics[width=0.45\linewidth]{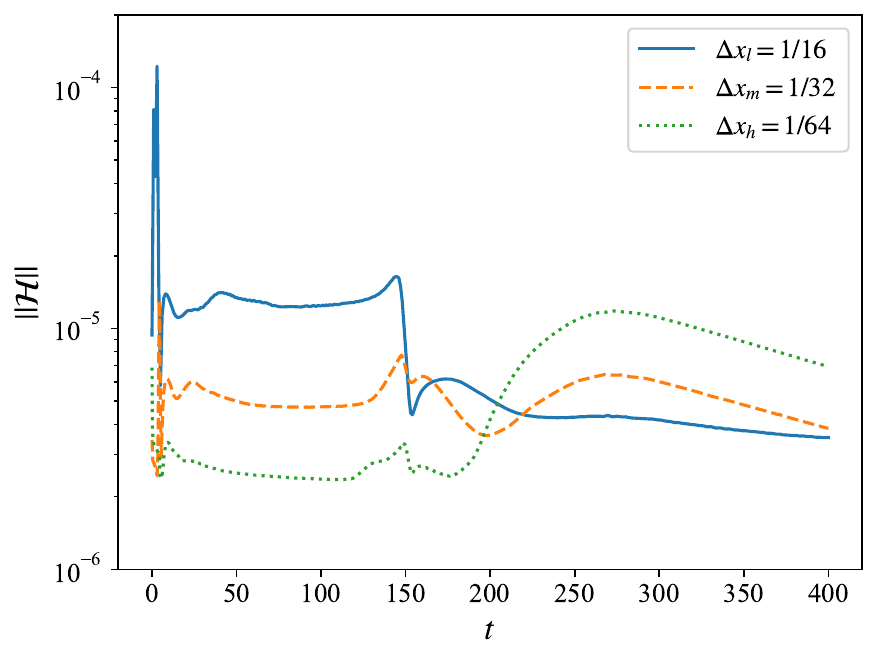}
    \includegraphics[width=0.45\linewidth]{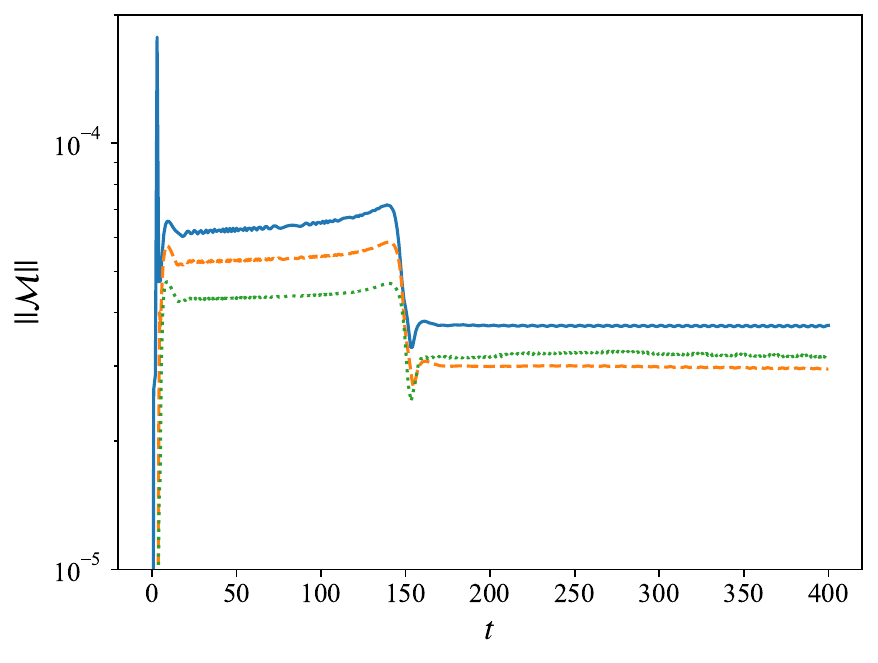}
    \caption{$L_2$ norm of the Hamiltonian constraint violation (left panel) and the magnitude of the momentum-constraint violation (right panel) as functions of time for the run \texttt{BSBH-q1-H} with $d = 40$ at three different resolutions.}
    \label{fig:convergence_constraints}
\end{figure*}

We use the run \texttt{BSBH-q1-H} with $d = 40$ to perform convergence tests of our numerical simulations. We use three different resolutions with finest grid spacings of $\Delta x_l = 1/16$, $\Delta x_m = 1/32$, and $\Delta x_h = 1/64$, which we refer to as low, medium, and high resolutions, respectively. The convergence order is estimated by comparing the differences between results at different resolutions. The $p$th-order convergence factor is defined as $Q_p \equiv (\Delta x_m^p - \Delta x_l^p) / (\Delta x_h^p - \Delta x_m^p)$. In Fig.~\ref{fig:convergence_GW}, we show the GW-energy differences between the different resolutions as a function of time. The differences between the high- and medium-resolution results are multiplied by the convergence factor $Q_p$ for $p = 2$ and $p = 3$. Comparing these with the differences between the low- and medium-resolution results, we find that the convergence order lies between 2 and 3. In Fig.~\ref{fig:convergence_constraints}, we show the $L_2$ norm of the Hamiltonian constraint violation (left panel) and the magnitude of the momentum-constraint violation (right panel) as functions of time for the three resolutions. We find that the constraint violations decrease with increasing resolution, further confirming the convergence of our simulations.

\begin{figure*}[htbp]
    \centering
    \includegraphics[width=0.98\linewidth]{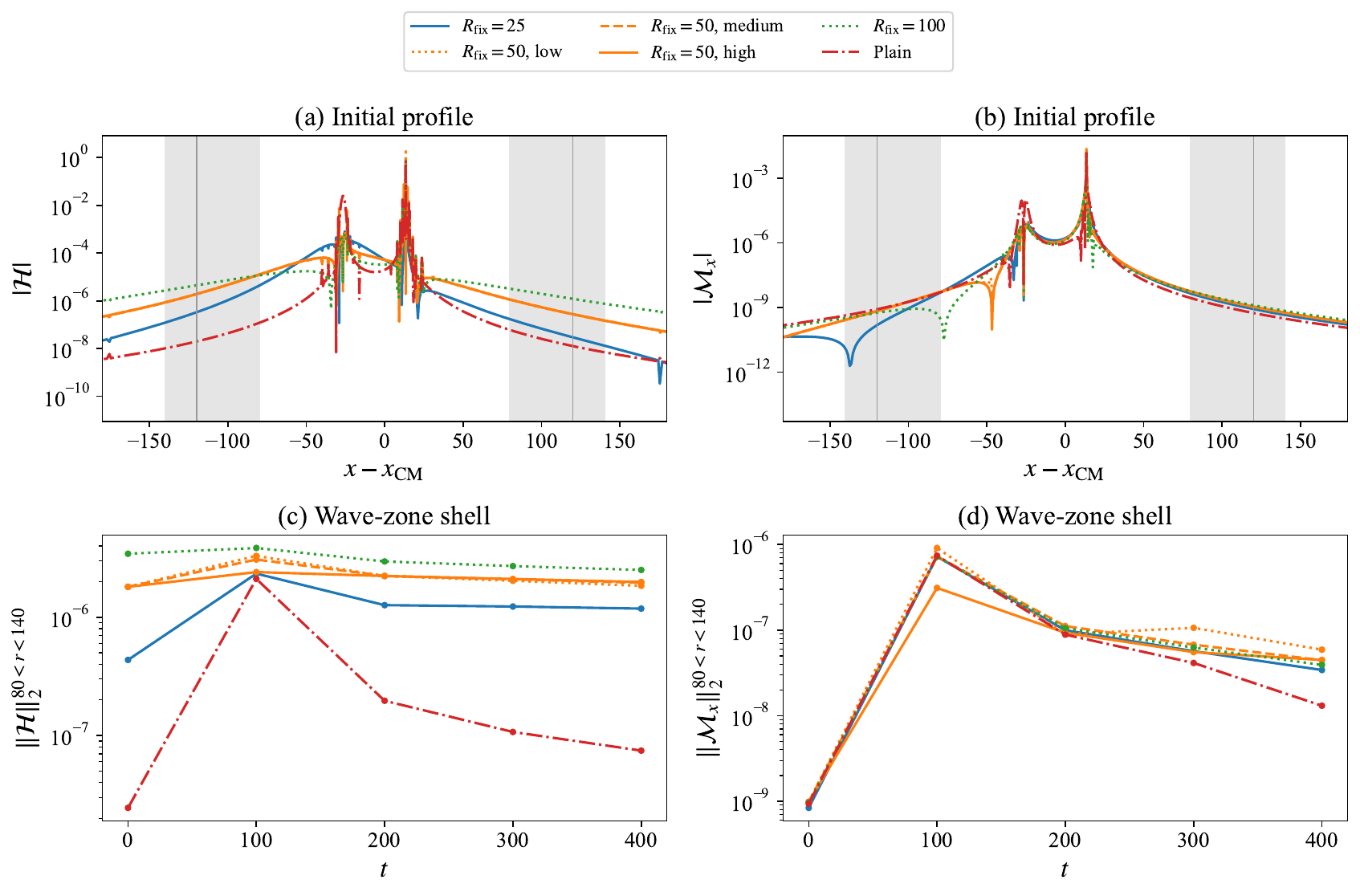}
    \caption{Wave-zone constraints for the $q = 0.5$ runs with three correction scales and plain superposition, including three resolutions for $R_\mathrm{fix} = 50$. The upper panels show the initial profiles; shaded regions denote $80 < |x - x_\mathrm{CM}| < 140$, and vertical lines denote $R_\mathrm{ex} = 120$. The lower panels show the coordinate-volume-weighted $L_2$ norms in the three-dimensional shell $80 < r < 140$.}
    \label{fig:wave_constraints}
\end{figure*}

We also measure the constraints directly in the extraction shell $\mathcal W = \{\bm{x}\in\Sigma_t:80 < |\bm{x}-\bm{x}_\mathrm{CM}| < 140\}$ for the \texttt{BSBH-q05} configuration. We use the coordinate-volume-weighted norm
\begin{equation}
    \|\mathcal H\|_{2}^{80 < r < 140}
    = \left[\frac{\int_{\mathcal W}|\mathcal H|^2\,\mathrm{d}^3x}
    {\int_{\mathcal W}\mathrm{d}^3x}\right]^{1/2},
\end{equation}
with an analogous definition for $\mathcal M_x$. Figure~\ref{fig:wave_constraints} shows the initial profiles and the shell norms during the evolution. The initial Hamiltonian norms for $R_\mathrm{fix} = 25$, $50$, and $100$ are $4.35\times10^{-7}$, $1.80\times10^{-6}$, and $3.44\times10^{-6}$, compared with $2.45\times10^{-8}$ for plain superposition. Thus the correction's long spatial tail has a measurable far-zone cost even as it improves the BS core. For the fiducial $R_\mathrm{fix} = 50$ data, the initial norm is essentially resolution independent, indicating an initial-data residual. During passage of the outgoing constraint pulse at $t = 100$, increasing the resolution lowers $\|\mathcal H\|_{2}^{80 < r < 140}$ from $3.29\times10^{-6}$ to $2.40\times10^{-6}$ and lowers $\|\mathcal M_x\|_{2}^{80 < r < 140}$ from $9.13\times10^{-7}$ to $3.12\times10^{-7}$.

After $t = 0$, the fiducial Hamiltonian norm remains between $1.95\times10^{-6}$ and $3.07\times10^{-6}$ through $t = 400$, and its largest pointwise value in the shell is $1.40\times10^{-5}$ as the pulse crosses. The $\sim10^{-4}$ level in the global convergence diagnostic is therefore dominated by the strong-field region and does not characterize the extraction zone. Moreover, the constraint pulse crosses the shell near $t \simeq 100$, whereas the merger waveform peaks near $t - R_\mathrm{ex} \simeq 160$, corresponding to extraction-coordinate time $t \simeq 280$ at $R_\mathrm{ex} = 120$. Although these tests cannot exclude every effect of the residual violations, the wave-zone residual is about two orders of magnitude below the global strong-field norm and is temporally separated from the merger waveform.

\section{Apparent-horizon matter diagnostics} \label{app:matter_diagnostics}

\begin{figure*}[htbp]
    \centering
    \includegraphics[width=0.9\linewidth]{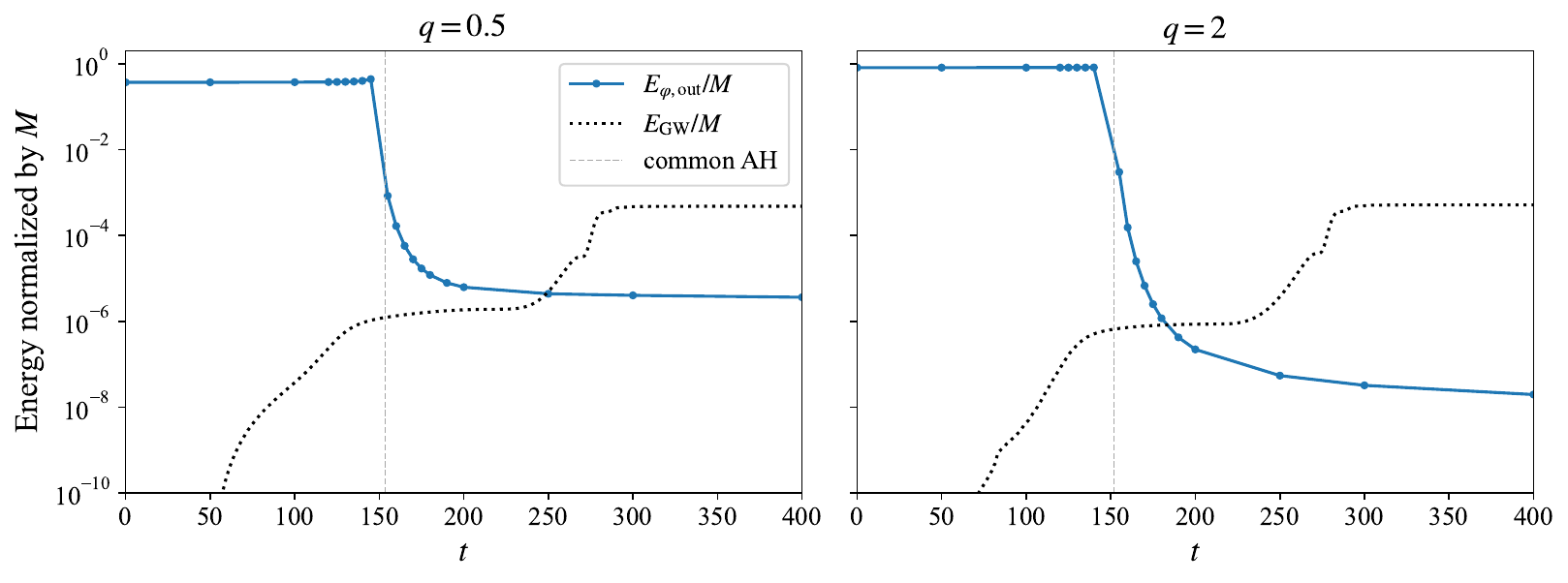}
    \caption{Scalar energy outside the apparent horizon and cumulative GW energy, normalized by the boosted initial binary mass $M$. The former is an Eulerian slice matter-energy integral and the latter is extracted at $R_\mathrm{ex} = 120$. Vertical dashed lines mark the first common AH.}
    \label{fig:energy_channels}
\end{figure*}

To separate the scalar matter remaining outside the BH from matter that has crossed the horizon, we perform dedicated $q = 0.5$ and $q = 2$ evolutions with the apparent-horizon (AH) finder enabled. Both configurations have boosted initial binary mass $M = 1.07914$. On each slice $\Sigma_t$, let $\Omega_\mathrm{out}(t)$ be the region exterior to the outermost physical AH. We define the exterior scalar energy by
\begin{equation}
    E_{\varphi,\mathrm{out}}(t)
    = \int_{\Omega_\mathrm{out}(t)}\rho\sqrt{\gamma}\,\mathrm{d}^3x.
    \label{eq:scalar-energy-outside-ah}
\end{equation}
Figure~\ref{fig:energy_channels} compares this Eulerian slice matter-energy integral with the cumulative GW energy extracted at $R_\mathrm{ex} = 120$. For $q = 0.5$, $E_\varphi(0)/M = 0.3691$ and the final $E_\mathrm{GW}/M = 4.791\times10^{-4}$; for $q = 2$, the corresponding values are $0.8131$ and $5.134\times10^{-4}$. At $t = 160$, $E_{\varphi,\mathrm{out}}/M$ is $1.66\times10^{-4}$ and $1.54\times10^{-4}$ for $q = 0.5$ and $q = 2$, respectively. By $t = 400$, these fractions fall to $3.63\times10^{-6}$ and $1.97\times10^{-8}$. Thus, in these two dedicated runs, the $q = 2$ case has a larger radiated GW energy and less surviving exterior scalar energy.

The quantity in Eq.~\eqref{eq:scalar-energy-outside-ah} contains the scalar rest, kinetic, gradient, and potential contributions measured by the Eulerian observer, but it does not subtract the negative gravitational binding energy as a separate additive term. The AH mass is a quasi-local horizon quantity, while the nonradiative gravitational field and interaction energy are not included in $M_\mathrm{AH} + E_{\varphi,\mathrm{out}} + E_\mathrm{GW}$. Because a dynamical spacetime has no gauge-invariant local gravitational-energy density, Fig.~\ref{fig:energy_channels} is a quantitative comparison of resolved channels rather than a closed additive budget. In particular, we do not assign all energy not carried by GWs uniquely to internal scalar excitations.

The total grid-integrated Noether charge does not by itself separate accretion from outer-boundary flux or numerical loss. We therefore formulate a finite-volume balance. Let $\Omega_{140}(t)$ be the region outside the outermost physical AH and inside a fixed coordinate sphere of radius $140$ centered at $\bm{x}_\mathrm{CM}$, and define
\begin{equation}
    N_\mathrm{out}^{(<140)}(t)
    = \int_{\Omega_{140}(t)}Q\,\mathrm{d}^3x,
    \quad
    Q \equiv \sqrt{-g}\,J^0.
\end{equation}
We also define $N_\mathrm{cloud}^{(<R)}$ as the charge outside the AH but inside an AH-centered sphere of radius $R$. This finite-aperture proxy tests whether surviving charge is localized near the BH; it is a subset of $N_\mathrm{out}^{(<140)}$ rather than a separate term in the balance.

Writing $\mathcal F^i = \sqrt{-g}J^i$, the charge flux into a moving horizon parametrized by $X^i(t,\theta,\phi)$ is
\begin{equation}
    \dot N_\mathrm{AH}
    = -\int_\mathrm{AH}\left(\mathcal F^i-Q\dot X^i\right)\mathrm{d}\Sigma_i,
    \label{eq:noether-ah-flux}
\end{equation}
where $\mathrm{d}\Sigma_i$ points out of the horizon interior, so positive $\dot N_\mathrm{AH}$ denotes accretion. The outward flux through the fixed sphere is
\begin{equation}
    \dot N_{140}
    = 140^2\int_0^{2\pi}\mathrm{d}\phi\int_0^\pi\mathrm{d}\theta\,
    \sin\theta\,\hat r_i\mathcal F^i
    \bigl(t,\bm{x}_\mathrm{CM}+140\hat{\bm r}\bigr),
\end{equation}
where positive $\dot N_{140}$ denotes charge leaving the control volume. Starting from $t_\star = 155$, when the outer common horizon has settled in both runs, we define the signed finite-volume residual
\begin{equation}
    \begin{aligned}
        \delta N_\mathrm{cons}^{(<140)}(t)
        &= N_\mathrm{out}^{(<140)}(t_\star)-N_\mathrm{out}^{(<140)}(t) \\
        &\quad -\int_{t_\star}^{t}\left(\dot N_\mathrm{AH}+\dot N_{140}\right)\mathrm{d}t.
    \end{aligned}
\end{equation}
The moving-boundary term in Eq.~\eqref{eq:noether-ah-flux} prevents charge advected toward the puncture from being misidentified as numerical loss.

Before contact, essentially all physical charge remains on the grid. We therefore define the empirical precontact conservation floor
\begin{equation}
    \epsilon_N^\mathrm{pre}
    = \max_{t \leq 120}\left|N_\mathrm{grid}(t)/N_0 - 1\right|,
\end{equation}
which is $2.61\times10^{-5}$ for $q = 0.5$ and $2.87\times10^{-6}$ for $q = 2$. The conservative numerical allowance and accreted lower limit are
\begin{align}
    \epsilon_N
    &= \max\left[
        |\delta N_\mathrm{cons}^{(<140)}(400)|/N_0,
        \epsilon_N^\mathrm{pre}
    \right], \\
    f_\mathrm{acc}^\mathrm{min}
    &= 1 - \frac{N_\mathrm{out}^{(<140)}(400)}{N_0}
       - \frac{N_{R=140}}{N_0} - \epsilon_N,
\end{align}
where $N_{R=140}$ is the charge escaping through the outer sphere over the full evolution; the initial charge outside this sphere is negligible at the quoted precision. The resulting fractions are listed in Table~\ref{tab:noether-accounting}. Before applying the numerical allowance, the nominal accreted fraction exceeds $99.999\%$ in both runs. Conservatively, we quote only $f_\mathrm{acc} > 99.99\%$.

\begin{table*}[htbp]
    \centering
    \caption{Noether-charge accounting at $t = 400$ for the dedicated AH evolutions. The cloud proxy uses an AH-centered aperture of radius $3r_{99}$.}
    \label{tab:noether-accounting}
    \begin{ruledtabular}
        \begin{tabular}{ccccccc}
            $q$ & $N_\mathrm{out}^{(<140)}/N_0$ & $N_\mathrm{cloud}^{(<3r_{99})}/N_0$ & $N_{R=140}/N_0$ & $\delta N_\mathrm{cons}^{(<140)}/N_0$ & $\epsilon_N$ & $f_\mathrm{acc}^\mathrm{min}$ \\
            $0.5$ & $9.20\times10^{-6}$ & $7.24\times10^{-7}$ & $1.49\times10^{-7}$ & $8.64\times10^{-6}$ & $2.61\times10^{-5}$ & $>99.99\%$ \\
            $2$ & $1.76\times10^{-8}$ & $2.30\times10^{-9}$ & $1.23\times10^{-9}$ & $1.02\times10^{-5}$ & $1.02\times10^{-5}$ & $>99.99\%$
        \end{tabular}
    \end{ruledtabular}
\end{table*}

\begin{figure*}[htbp]
    \centering
    \includegraphics[width=0.9\linewidth]{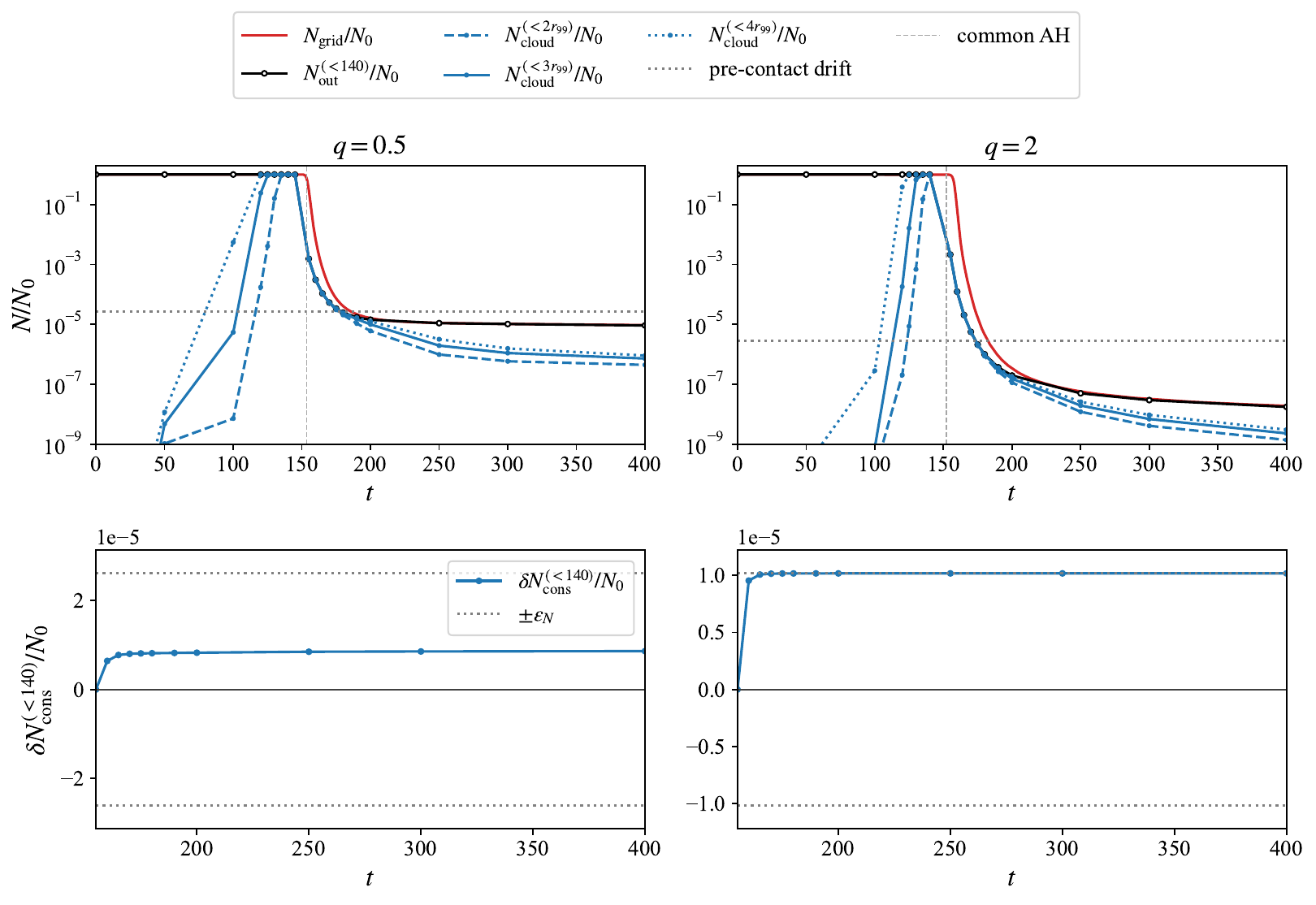}
    \caption{Noether-charge accounting for the dedicated $q = 0.5$ and $q = 2$ evolutions. The upper panels show the grid-integrated charge, the charge outside the physical AH and inside the fixed $R = 140$ sphere, and AH-centered cloud proxies. The lower panels show the signed finite-control-volume residual from $t_\star = 155$; dotted lines denote the conservative allowance $\epsilon_N$.}
    \label{fig:noether-accounting}
\end{figure*}

Figure~\ref{fig:noether-accounting} shows the charge decomposition and the time-dependent balance residual. The cloud proxies grow before contact as the BS enters the AH-centered aperture and then fall to $10^{-7}$--$10^{-9}$ of the initial charge. The finite-volume residual remains at the few-$10^{-5}$ level. We cannot isolate a nonzero numerical-loss contribution below this diagnostic accuracy, and the late-time cloud values are also below these conservative allowances. They are therefore upper limits rather than detections of a gravitational atom.

\section{The $(3,0)$ mode as a numerical cautionary example}

For completeness, we show the numerical-sensitivity tests of the $(3,0)$ mode in this appendix. The tests do not establish the $(3,0)$ component as a prescription-independent physical signal. We present them solely as a cautionary example of the sensitivity of subdominant waveforms to numerical resolution and initial-data construction. With the present calculations, we cannot distinguish among a numerical artifact, a prescription-dependent feature, and a genuine physical signal.

\begin{figure*}[htbp]
    \centering
    \includegraphics[width=0.98\linewidth]{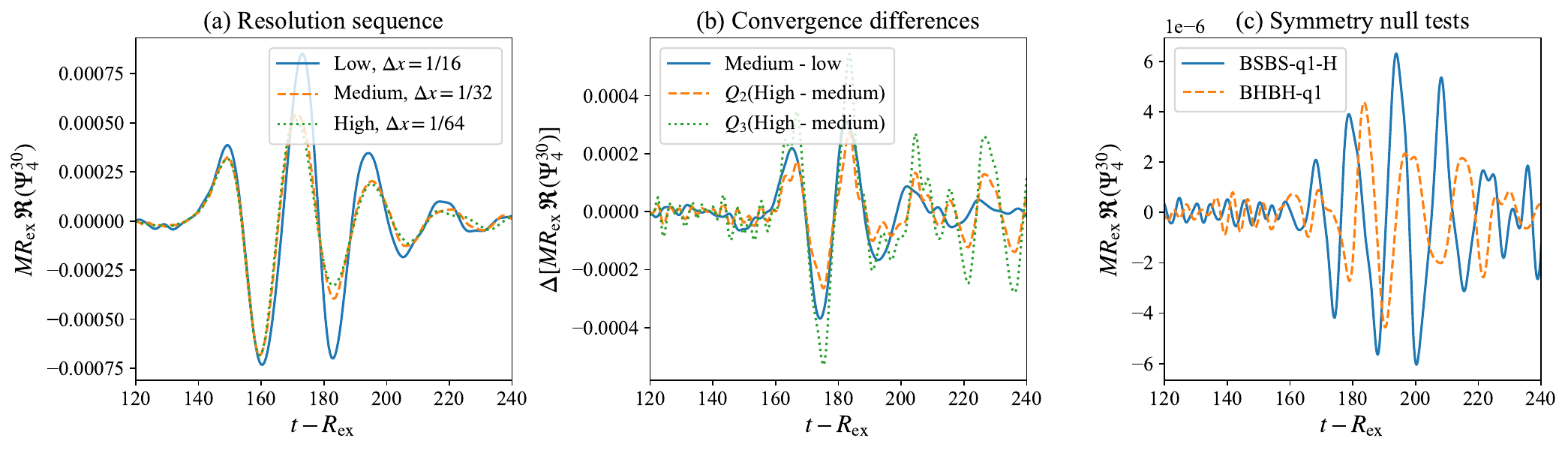}
    \caption{Resolution and symmetry null tests for the $(3,0)$ mode at $R_\mathrm{ex} = 120$. Panels (a) and (b) show the three-resolution sequence for \texttt{BSBH-q1-H} and the corresponding scaled differences. Panel (c) shows the symmetry-forbidden residuals in the equal-mass \texttt{BSBS-q1-H} and \texttt{BHBH-q1} simulations at the fiducial resolution.}
    \label{fig:GW30_q1_tests}
\end{figure*}

First, we perform a $(3,0)$ resolution test for the equal-mass \texttt{BSBH-q1-H} configuration. As shown in Figs.~\ref{fig:GW30_q1_tests}(a) and~\ref{fig:GW30_q1_tests}(b), the peak amplitudes at low, medium, and high resolution are $8.5275\times10^{-4}$, $6.8444\times10^{-4}$, and $6.8351\times10^{-4}$. The low-resolution signal has a visible timing and amplitude error, whereas the medium- and high-resolution peaks differ by $0.14\%$ and occur at the same sampled time. We also use the equal-mass BS-BS and BH-BH runs as symmetry null tests. Reflection symmetry about the center of mass forbids their odd-$\ell$ signals in the collision-axis frame. Figure~\ref{fig:GW30_q1_tests}(c) shows peak numerical residuals of $6.31\times10^{-6}$ for BS-BS and $4.56\times10^{-6}$ for BH-BH, corresponding to $0.92\%$ and $0.67\%$ of the $q = 1$ BS-BH peak. The extraction and rotation procedure therefore does not by itself generate an odd-$\ell$ signal at the observed BS-BH amplitude. This null test addresses only one possible source of contamination and is not evidence that the BS-BH component is a robust physical signal.

\begin{figure*}[htbp]
    \centering
    \includegraphics[width=0.98\linewidth]{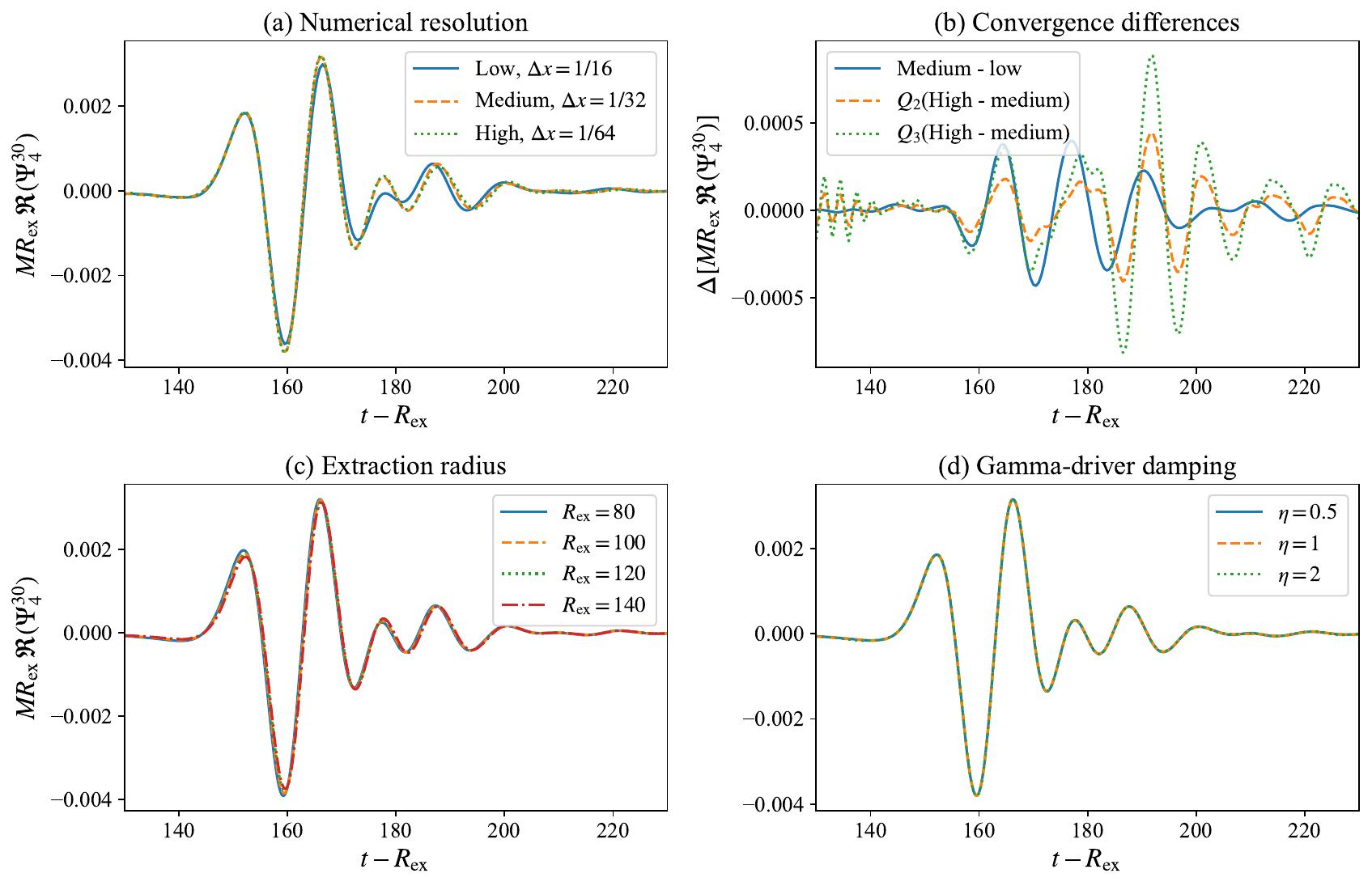}
    \caption{Sensitivity tests for the $(3,0)$ mode of the $q = 0.5$ BS-BH collision. Panels (a) and (b) show the three-resolution sequence and the scaled differences. Panels (c) and (d) show the extraction-radius and Gamma-driver-damping dependence, respectively.}
    \label{fig:GW30_robustness}
\end{figure*}

Second, we perform dedicated sensitivity tests for the unequal-mass $q = 0.5$ configuration. Figure~\ref{fig:GW30_robustness} compares the waveform at three resolutions, the corresponding scaled differences, four extraction radii, and three Gamma-driver damping parameters. At $R_\mathrm{ex} = 120$, the peak values of $|M R_\mathrm{ex}\Re(\Psi_4^{30})|$ at low, medium, and high resolution are $3.6221\times10^{-3}$, $3.8042\times10^{-3}$, and $3.8267\times10^{-3}$, respectively. The peak changes by $5.03\%$ from low to medium resolution and by $0.59\%$ from medium to high resolution, while the peak time is unchanged at the available sampling interval of $0.25$.

Varying the extraction radius from $R_\mathrm{ex} = 80$ to $140$ changes the finite-radius peak by $4.5\%$, while the retarded peak times agree within $0.25$. Changing the Gamma-driver damping over $\eta \in \{0.5,1,2\}$ changes the peak by only $0.15\%$.

\begin{figure*}[htbp]
    \centering
    \includegraphics[width=0.98\linewidth]{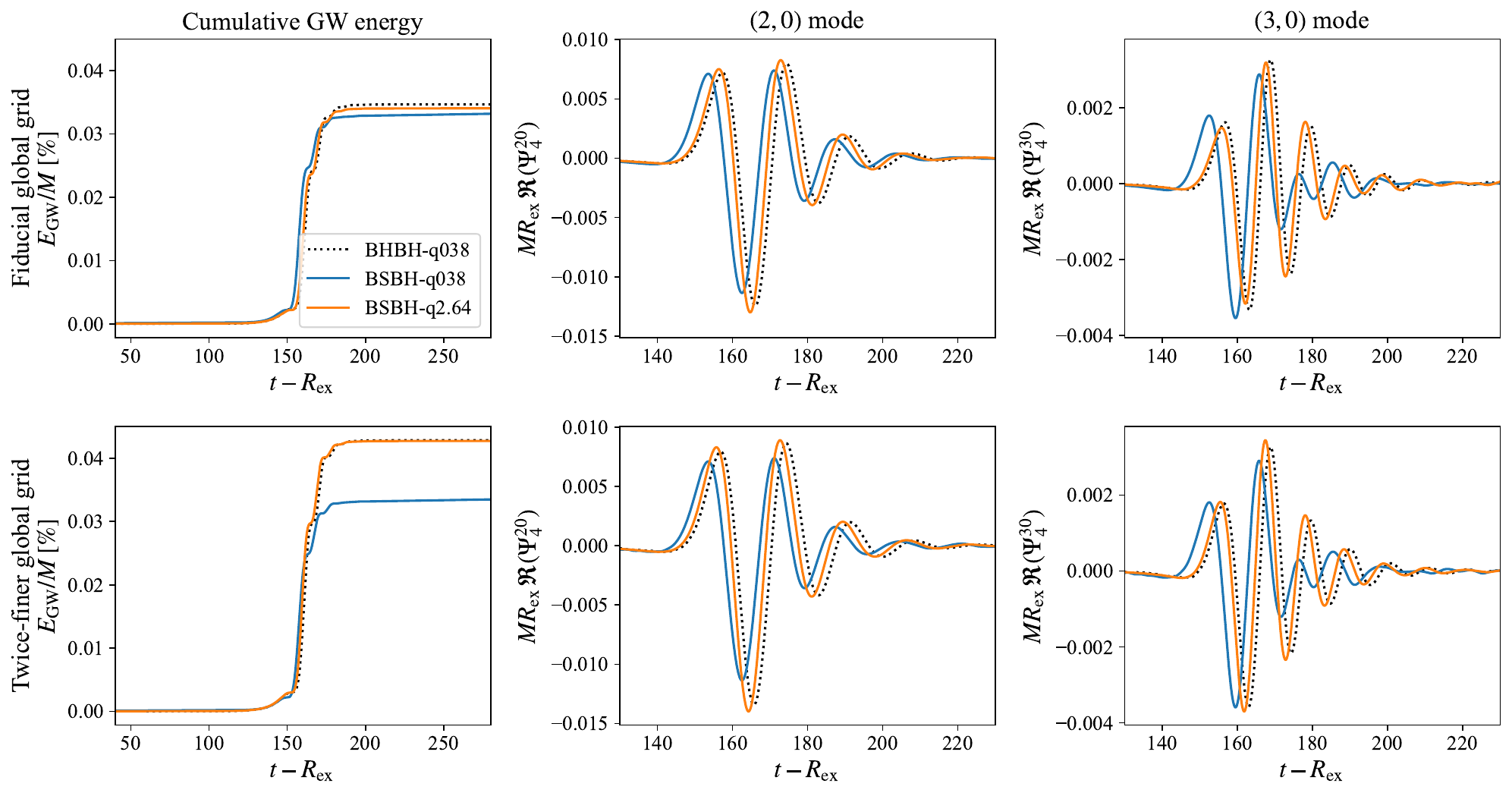}
    \caption{Cumulative GW energy and the $(2,0)$ and $(3,0)$ modes for the reciprocal $q = 0.38$ and $q = 2.64$ BS-BH configurations and the matched BH-BH baseline, on the fiducial and twice-finer global grids. Runs containing the $M_\mathrm{BH} = 0.2700$ puncture use one additional AMR level.}
    \label{fig:q038_high_res}
\end{figure*}

Third, the configurations containing the small $M_\mathrm{BH} = 0.2700$ puncture require a separate check. In addition to one extra AMR level, we evolve \texttt{BSBH-q2.64}, \texttt{BSBH-q038}, and \texttt{BHBH-q038} with base grids of $N = 256$ and $512$, denoted by fiducial and twice-finer global grids in Fig.~\ref{fig:q038_high_res}. The $q = 0.38$ BS-BH radiated energy changes by $0.87\%$, whereas the $q = 2.64$ BS-BH and shared BH-BH energies change by $25.4\%$ and $23.5\%$, respectively. The ratio $E_\mathrm{GW}^{q=2.64}/E_\mathrm{GW}^\mathrm{BHBH}$ changes from $0.98225$ to $0.99725$, so we use the twice-finer results in Fig.~\ref{fig:unequal} and regard the two energies as comparable at the highest available resolution.

For the $q = 0.38$ $(3,0)$ signal relative to its matched BH-BH baseline, the relative peak-time offset changes from $3.75$ to $3.50$ between the two grids. Since the time series is sampled at intervals of $0.25$, we assign a resolution uncertainty of $0.25$ to this timing difference. The corresponding peak-amplitude ratio changes from $1.066$ to $1.003$, so an amplitude enhancement is not resolved. This test shows that the $(3,0)$ mode difference is not robust under global refinement.

\begin{figure*}[htbp]
    \centering
    \includegraphics[width=0.98\linewidth]{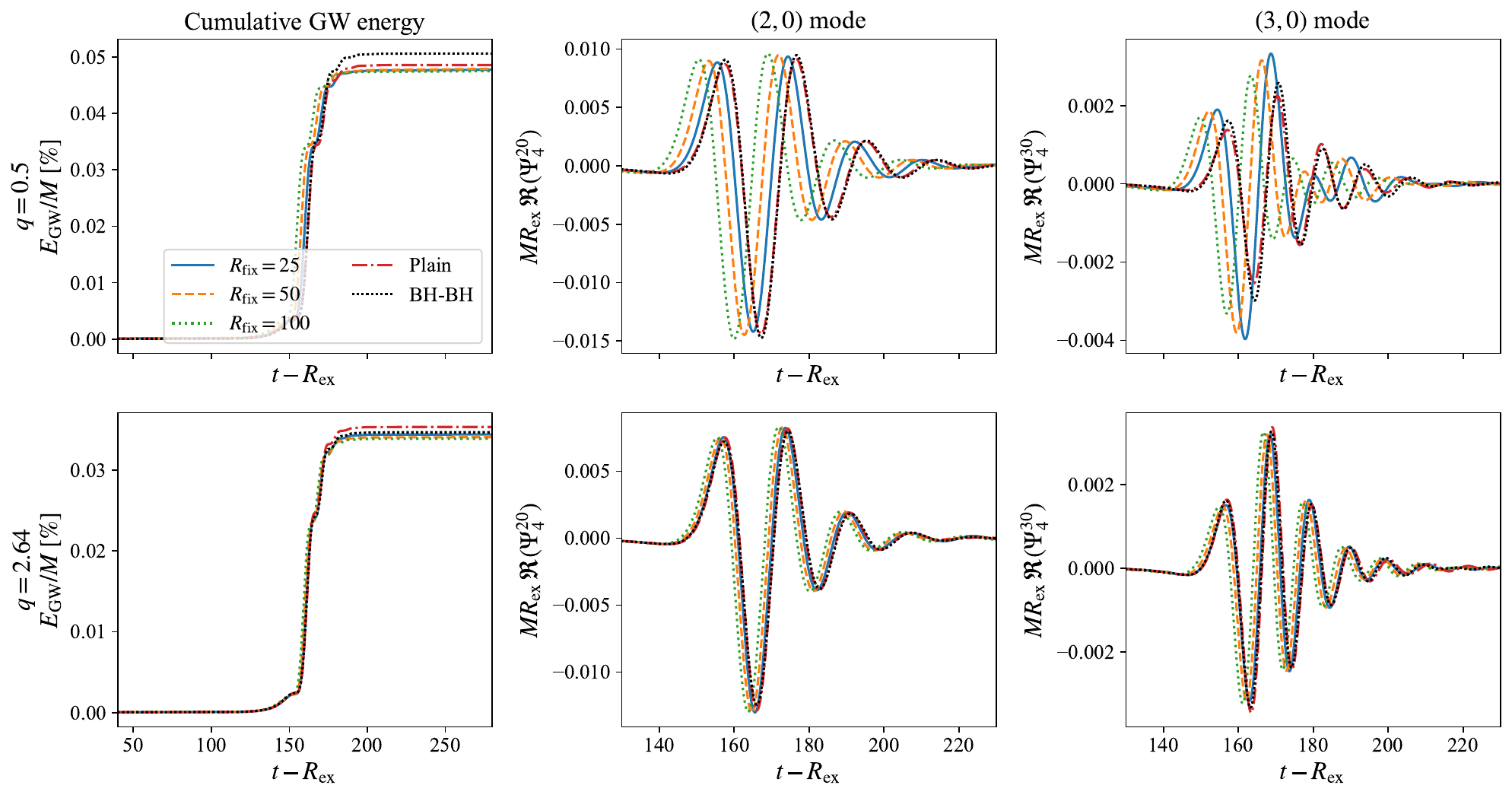}
    \caption{Dependence of the cumulative GW energy and the $(2,0)$ and $(3,0)$ modes on the correction scale and on plain superposition for $q = 0.5$ and $q = 2.64$. The matched BH-BH results are included as references.}
    \label{fig:GW30_id_dependence}
\end{figure*}

We next test the dependence on the one-body conformal correction by varying $R_\mathrm{fix}$ over $25$, $50$, and $100$ and comparing with plain superposition for $q = 0.5$ and $q = 2.64$. Figure~\ref{fig:GW30_id_dependence} shows that the total energy and dominant quadrupole are comparatively insensitive to the corrected profile. The corrected $q = 0.5$ energies span $(5.126$--$5.170)\times10^{-4}$, a $0.86\%$ range, while the principal $(2,0)$ difference is a timing shift. For $q = 2.64$, the corrected energies span $(3.344$--$3.396)\times10^{-4}$. Plain superposition differs somewhat more but retains the same dominant-mode morphology.

The subdominant mode is more sensitive to the correction profile. For $q = 0.5$, the $(3,0)$ peaks for $R_\mathrm{fix} = 25$, $50$, and $100$ exceed the matched BH-BH peak by approximately $33\%$, $28\%$, and $11\%$, respectively, while their relative peak-time offsets range from $2.5$ to $7.25$. Plain superposition shows neither an amplitude enhancement nor a peak-time shift, although it is not an equally reliable description of the BS bulk because it produces the core-volume distortion and premature collapse demonstrated in Sec.~\ref{sec:results}. For $q = 2.64$, the $(3,0)$ peak is less sensitive within the corrected family, but its timing still varies. This strong prescription dependence prevents us from assigning physical significance to the $(3,0)$ mode calculated here.

Taken together, these results do not satisfy the combined robustness standard needed to identify the $(3,0)$ component as a physical signal. In particular, the $q = 0.38$ amplitude difference changes under global refinement, while the $q = 0.5$ amplitude and timing depend strongly on the freely chosen correction scale. Independent confirmation with alternative or fully constraint-solved initial data would be required before assigning a physical interpretation to this mode.

\bibliography{ref.bib}

\end{document}